\documentclass{article}

\usepackage[utf8]{inputenc}
\usepackage{graphicx}
\usepackage{natbib}
\usepackage{hyperref}
\usepackage{xr}

\usepackage[english]{babel}
\usepackage{float}
\usepackage{amsmath}
\usepackage{fullpage}
\usepackage{caption}
\usepackage{ifthen}
\usepackage[format=hang]{subfig}
\usepackage{threeparttablex}
\usepackage{tabularx}
\usepackage{longtable}
\usepackage{booktabs}
\usepackage{geometry}
\usepackage{lscape}
\usepackage{afterpage}
\usepackage{xcolor}
\usepackage{placeins}
\usepackage{multicol}
\usepackage{supertabular}
\usepackage{amssymb}
\usepackage{supertabular}
\usepackage{array}

\setcitestyle{authoryear,open={(},close={)}} 
\setcitestyle{citesep={;}}

\definecolor{magenta_custom}{rgb}{1, 0, 1}
\definecolor{green_custom}{rgb}{0, 0.7, 0}

\graphicspath{ {images/} }

\begin{document}

\title{
    An updated view of Jupiter's UV main auroral emission: large-scale analysis of UV and particle data from Juno's first 60 orbits
}
\author{
    L. A. Head \thanks{Corresponding author: LA.Head@uliege.be (email address)},
    D. Grodent,
    B. Bonfond,
    N. Kruegler,
    A. Sulaiman,\\
    G. Clark,
    B. Benmahi,
    G. Sicorello,
    J. Vinesse,
    T. K. Greathouse,\\
    R. Giles,
    B. H. Mauk
}
\date{\today}

\abstract{
     The properties and origins of Jupiter's UV main auroral emission are, at present, not yet fully understood, particularly the extent to which its two distinct zones, the Zone-I and Zone-II aurora, originate from a common source. 
     In this work, remote-sensing and in-situ data from the Juno spacecraft are used to investigate and produce maps of the average properties of the main emission over Juno's first 60 orbits. 
     It is found that the peak in mean energy of electrons precipitating in the main emission is consistently poleward of the energy-flux (or brightness) peak, in both hemispheres and at all local times, and is co-located with the statistical location of the Zone-II aurora.
     Evidence is presented that the flux-energy relationship of the main emission is adequately characterised by theoretical power laws describing electron acceleration from quasi-static potentials or Alfvénic turbulence and that it shows a modulation with ionospheric electron density consistent with theoretical predictions.
     Finally, the average energy spectra of precipitating electrons in the Zone-I aurora remain power-law-like between an energy flux of 1 and 100 mW m$^{-2}$, whereas the average Zone-II spectra develop a distinct ``plateau'' or ``flattop'' shape beyond an energy flux of $\sim$5 mW m$^{-2}$.
     Our results provide some evidence that the Zone-I and Zone-II aurorae are phenomenologically distinct, and the consequences of this interpretation for future investigation of the main emission are discussed.
     
}

\maketitle

\section{Introduction}
\label{sec:introduction}

The main emission of Jupiter's ultraviolet (UV) aurora is a permanent, semi-continuous band of auroral emission at all magnetic longitudes.
Its origins are the subject of continuous study and debate within the field.
At present, two leading theories exist.
The first is the ``current-based'' theory, in which loops of electrical current in Jupiter's magnetosphere that arise from the breakdown of corotation of magnetospheric plasma provoke quasi-static electrical potentials above the ionosphere that drive electrons into the atmosphere and produce the main emission \citep{cowley+bunce:2001,nichols:2011}.
The second is the ``Alfvénic'' or ``turbulence'' theory, in which plasma turbulence in the equatorial plasma sheet of the middle magnetosphere launches Alfvén waves along magnetic-field lines toward the ionosphere which interact with counter-propagating reflected Alfvén waves to cascade to electron-inertial wavelengths, which are then able to provoke meaningful electron acceleration in the ionosphere and hence the main emission \citep{saur+:2003,saur+:2018}.
Although the current-based theory was broadly accepted in the community prior to the Juno mission, the relative lack of discrete ``inverted-V'' electron-energy signatures, taken as a sign of acceleration via quasi-static potentials, and relative abundance of bidirectional broadband electron distributions above the aurora \citep{mauk+:2017_mainemission,salveter+:2022} has led to increased debate about the relative contribution of these two acceleration mechanisms to the main emission \citep{bonfond+:2020,sulaiman+:2022,provan+:2026}, though both are expected to contribute to some extent \citep[e.g.][]{feng+:2022} and may be related, with Alfvén waves generated at low altitude rather than in the magnetodisc \citep{sulaiman+:2026}.

The energy flux and mean energy of precipitating electrons are key observational parameters in investigating the origins of the main emission.
Previous work using observations from the Hubble Space Telescope (HST) found a power-law relationship between mean precipitating electron energy and energy flux \citep{gustin+:2004,gerard+:2016}, which was later supported by in-situ observations from Juno in a limited number of cases \citep{clark+:2018}.
Both the current-based and Alfvénic theories predict power-law-like relationships between the energy flux and mean electron energy with similar exponents \citep{cowley+bunce:2001,saur+:2003,clark+:2018}.
Additionally, factors such as ionospheric electron density, which has been observed to be remarkably low in regions of Jupiter's aurora \citep{allegrini+:2021,kruegler+:2025}, are expected to influence the properties of the main emission, though this is yet to be observationally verified in a quantitative way.
Observations made by the JADE (Jovian Auroral Distributions Experiment; \citealt{mccomas+:2017}) instrument on board Juno have indicated that the average energy of precipitating electrons may increase poleward within the main emission \citep{allegrini+:2020}.
Besides the relationship between the energy flux and average energy of precipitating electrons, JADE data have also indicated that the energy distribution of precipitating electrons within the main emission may also depend on the total energy flux \citep{ebert+:2019}, transitioning from a power law to a broadband ``flattop'' distribution at higher energy fluxes.

Observations of directional energetic-particle fluxes and field-aligned currents above the main emission have allowed for the identification of two distinct zones: the Zone-I aurora, in which the field-aligned currents are largely upward (away from the planet) and is hence dominated by downward-travelling electrons; and the Zone-II aurora, in which the opposite is true \citep{mauk+:2020,sulaiman+:2022}.
Although these two zones are often contained within the main emission and are hence visually indistinguishable, under conditions of compression of the magnetosphere, the Zone-II aurora may become noticeably separated from the Zone-I aurora in the form of bright ``bridges'' just poleward of (the rest of) the main emission \citep{head+:2025}.
This is rather striking, since this means that the downward-current Zone-II aurora has comparable brightnesses to the rest of the main emission, indicating that, although the Zone-II is dominated by upward-travelling electrons, there remains a sufficient flux of downward electrons to provoke meaningful auroral emission \citep{mauk+:2020}.
The Zone-I and Zone-II aurora are both considered to be part of the main emission, though there is evidence to suggest that the two zones arise from distinct acceleration processes, or, at least, that various acceleration processes may occur to different extents along magnetic field lines threading the Zone-I and Zone-II aurora.
The most noteworthy example of this is the association between upward electron beams and the generation of electrostatic waves above the Zone-II aurora \citep{elliott+:2020,sulaiman+:2022,head+:2025}, though it is not entirely clear to what extent these features contribute to the generation of the Zone-II aurora rather than simply be coincident on Zone-II field lines or produce other non-Zone-II aurora poleward of the main emission.


Thus, this work aims to investigate to what extent the Zone-I and Zone-II aurorae are phenomenologically distinct, and whether these two zones contribute toward a single phenomenologically coherent main emission.
Additionally, we aim to determine the typical energy-flux and mean-energy profiles of these two zones and the main emission as a whole, and how factors such as ionospheric electron density may influence the properties of the main emission.
To this end, we derive statistical arguments from the properties of the aurora over Juno's first 60 perijoves using both remote-sensing and in-situ observations.

\section{Observations and methods}
\label{sec:observations}

Juno, currently in an elliptical polar orbit around Jupiter, is equipped with a number of instruments; this work uses data from the UltraViolet Spectrograph (UVS; \citealt{gladstone+:2017_uvs}), the Jupiter Energetic particle Detector Instrument (JEDI; \citealt{mauk+:2017_jedi}), and the FluxGate Magnetometer (FGM or MAG; \citealt{connerney+:2017}).
Data from UVS, which has a narrow field of view that captures ``strips'' of Jupiter's aurora during each of Juno's 30-second spins, can be combined to produce composite images or maps of the UV aurora.
This work uses a single composite UVS image of the aurora for every low-altitude pass of Juno over each hemisphere of the aurora during a perijove (PJ; e.g. PJ1-N for perijove 1, northern hemisphere).
These images are produced by collating data from the 100 consecutive Juno spins centred as close as possible to the perijove time while covering at least 75\% of the auroral region \citep{bonfond+:2018}.
Radiation noise from the impact of high-energy electrons on the detector is also removed \citep{bonfond+:2021}.
More detail regarding the production of these maps is given in \citet{head+:2024}.
The map in the 155-162 nm range, which is essentially unabsorbed by hydrocarbons, can be used to recover the total H$_2$ brightness in the UV range through multiplication by a factor of 8.1 \citep{gustin+:2012}.
\citet{vinesse+:2026} noted that this range is sometimes affected by a saturation effect in UVS data and that this effect can be detected and corrected by comparing this map to one acquired in the 163-165 nm range; this correction is used in this work. 
Furthermore, the ratio between the corrected 155-162 nm map and the 135-140 nm map, the latter wavelengths being absorbed by methane, can be combined with the elevation angle to derive the mean energy of the precipitating electrons.
It should be noted that such a conversion relies on a number of assumptions concerning the vertical composition profile of the atmosphere and the energy distribution of the precipitating particles.
Here, we used the Transplanet model \citep{benmahi+:2024}, assuming the vertical profile from \citet{grodent+:2001} and a kappa broadband electron energy distribution; the uncertainties associated with these choices are discussed in \citet{benmahi+:2024}.
Moreover, the total H$_2$ UV brightness can also be converted into a precipitating energy flux using 1 kR $\sim$ 10 mW m$^{-2}$ \citep{gustin+:2012}, as this conversion only mildly varies with the mean electron energy.  

Data from the in-situ instruments JEDI (via the JMIDL tool provided by the John Hopkins University; \citealt{mauk+:2017_jedi}) and MAG (via the AMDA repository and associated \texttt{speasy} Python library provided by the CDPP; \citealt{genot+:2021,jeandet+:2024}) are also used; more technical detail is given in \citet{head+:2025} and the relevant instrument references above.
JEDI data from its three detectors, which capture electron fluxes in the 20 - 1000 keV range, are combined to produce complete (though frequently missing the electrons in the most field-aligned channel <5\textdegree; \citealt{mauk+:2017}) energy spectra of precipitating electrons.
Lower-energy electrons below 20 keV are not detected by JEDI but are detected by the JADE instrument \citep{mccomas+:2017}; high-energy electrons above 100 keV are estimated to contribute 70-80\% of the total energy flux to the main emission \citep{allegrini+:2020} and so we consider JEDI measurements to constitute a broadly representative view of the properties of precipitating auroral electrons.
Additionally, calibration difficulties make electron measurements below 30 keV unreliable and so this energy range is ignored.
Further details regarding the response of JEDI to conditions in Jupiter's magnetosphere and the corrections that are applied are available in the JEDI User Guide \citep{mauk+:2026}.
Field-aligned currents can be calculated from MAG data compared against the latest JRM33 magnetic-field model \citep{connerney+:2022} implemented using the \texttt{JupiterMag} Python wrapper \citep{jupitermag,wilson+:2023} and extrapolated via magnetic-flux conservation to the assumed auroral altitude of 400 km following \citet{al_saati+:2022}; further detail regarding this calculation and its limitations is given in the appendix of \citet{head+:2025}.
The JRM33 magnetic-field model is also used to map Juno's location to its magnetically linked ``footprint'' in the aurora via field-line tracing.

To identify the boundaries of the main emission in each auroral image used, masks were produced using the methods presented by \citet{groulard+:2024}.
However, instead of using the statistical position of the main emission from \citet{bonfond+:2012} as used by \citet{groulard+:2024}, detected main-emission contours were produced for each auroral image using the methods presented by \citet{head+:2024}.
These contours formed the starting point for the mask-creation algorithm described by \citet{groulard+:2024}.
These masks were manually checked to ensure they provided a sensible fit to the visual morphology of the main emission.
\citet{groulard+:2024} provided two types of masks (one excluding and one including the ``bridge'' or parallel arcs features); more recent work has suggested that the bridges observed in the polar region are likely manifestations of the Zone-II of the main emission \citep{head+:2025} and thus we opted to use the broader main-emission masks that include these features.

Cases with obvious dawn storms (of which 11 were identified during this work), which are transient features with elevated electron energy fluxes and mean energies coincident with the dawn-side main emission \citep{bonfond+:2021, ebert+:2021}, have been removed from this analysis in order to isolate the properties of the main emission, which is the feature under investigation in this work.
Nevertheless, it is expected that smaller (pseudo) dawn storms \citep{bonfond+:2021} may have escaped this filtering and may influence the average properties of the dawn-side main emission; this is not expected to greatly affect the conclusions of this work nor the reliability with which we might draw them.


The main-emission crossings considered in this work are given in Table \ref{tab:crossing_table}.
Some main-emission crossings may not be present in Table \ref{tab:crossing_table} for a number of reasons - such as if they occurred with an altitude above 2 R$_{J}$ (above the expected main-emission acceleration altitude; \citealt{sulaiman+:2022}), occasional gaps in the data provided by AMDA, or simply that Juno did not cross the main emission when passing over the auroral region - which brings the total number of considered cases down from 120 (60 perijoves, two hemispheres) to 92.
From the current density calculated from MAG data, it is possible to identify the upward-current Zone-I and downward-current Zone-II of the main emission \citep{mauk+:2020}.
In this work, crossings of the Zone-I and Zone-II aurora were considered to be reliably identified when:
\begin{itemize}
	\item the Juno magnetic footprint was visually within or very close ($\lesssim$ 300 km) to the UV main emission in the energy-flux maps;
	\item the JEDI downward-electron energy spectra contained increased fluxes or distinct features not found in the ``background'' spectrum; and
	\item the current density determined from MAG data showed a clear ``up-down'' feature indicative of the Zone-I/II aurora \citep{mauk+:2020,al_saati+:2022}.
\end{itemize}
These crossing intervals were found to be in near-perfect agreement with crossings identified by \citet{kruegler+:2026}, though both this work and the work of \citet{kruegler+:2026} identified crossings not found by the other.
Crossings identified by \citet{kruegler+:2026}, where not already identified using the criteria above, have been included in this work, as we consider their own identification methods to be equally valid as the methods used in this work.
In all, this led to 47 reliable identifications of Zone-I/II crossings during the first 60 perijoves, of which 31 are within the first 30 perijoves; the details of these crossings are given in Appendix \ref{app:crossing_table}.
These identified crossings are biased toward earlier perijoves for two reasons.
Firstly, in the southern hemisphere, Juno crossings of the main emission occur at ever greater altitudes as the mission progresses, leading to greater uncertainty in the Juno-footprint mapping in the aurora, reduced current-density magnitudes, and masking of features in the electron-energy spectra by interference from Jupiter's radiation belts; this makes it more difficult for all three conditions to be fulfilled for later perijoves.
Secondly, in the northern hemisphere, toward later perijoves, Juno crosses the main emission at very low altitude, which increases the influence of inaccuracies in the JRM33 magnetic-field model and hence means that MAG measurements can greatly deviate from the expected values even in the absence of field-aligned currents, masking these latter features even when they are indeed present above the aurora.
It is naturally expected that the Zone-I and Zone-II aurora still exist in later perijoves but merely that they become increasingly challenging to identify on a case-by-case basis.

\section{Results}

\subsection{Analysis of the full auroral maps}
\label{sec:results_remote}

\begin{figure}[tb]
    \centering
    \includegraphics[width=\linewidth]{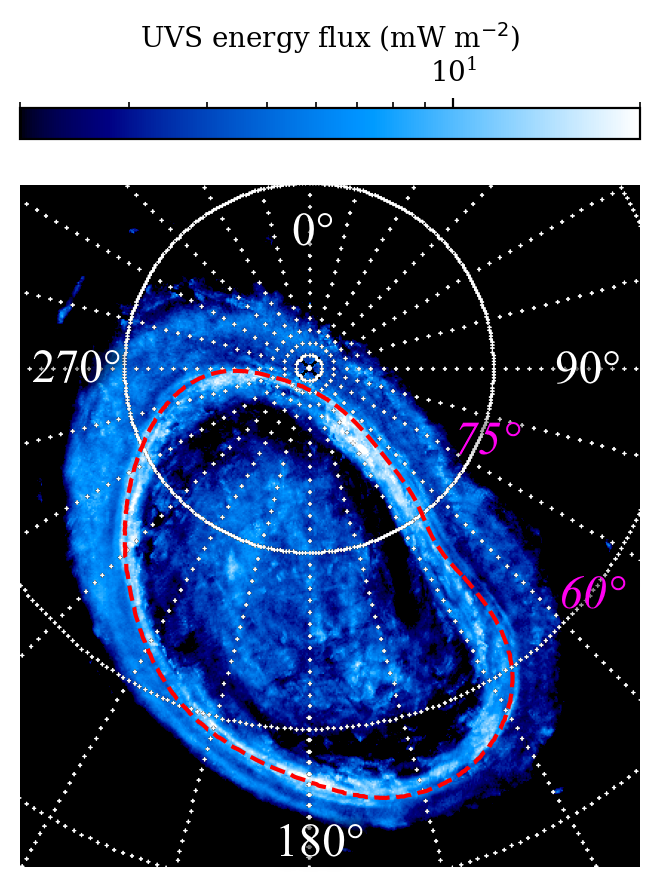}
    \caption{
        Stacked average map of auroral electron energy flux inferred from UVS in the northern hemisphere from PJ1 to PJ60. Gridlines denote planetographic latitude (magenta text) and System-III west longitude (white text) with a spacing of 15\textdegree. The average main-emission contour from \citet{head+:2024} is given by the dashed red line.
    }
    \label{fig:map_uv_flux_N}
\end{figure}
\begin{figure}[tb]
    \centering
    \includegraphics[width=\linewidth]{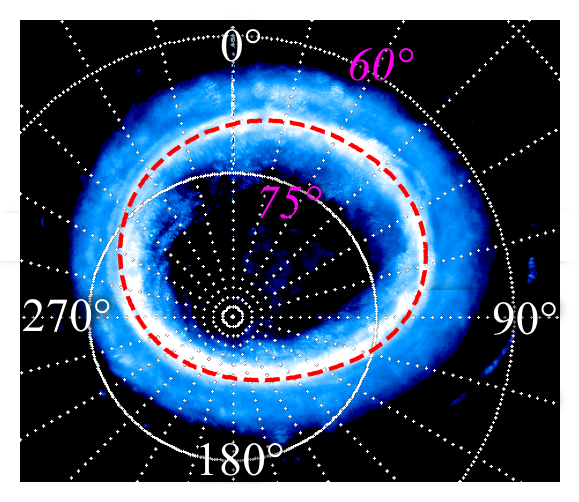}
    \caption{
        As Figure \ref{fig:map_uv_flux_N} but for the southern hemisphere.
    }
    \label{fig:map_uv_flux_S}
\end{figure}
\begin{figure}[bt]
    \centering
    \includegraphics[width=\linewidth]{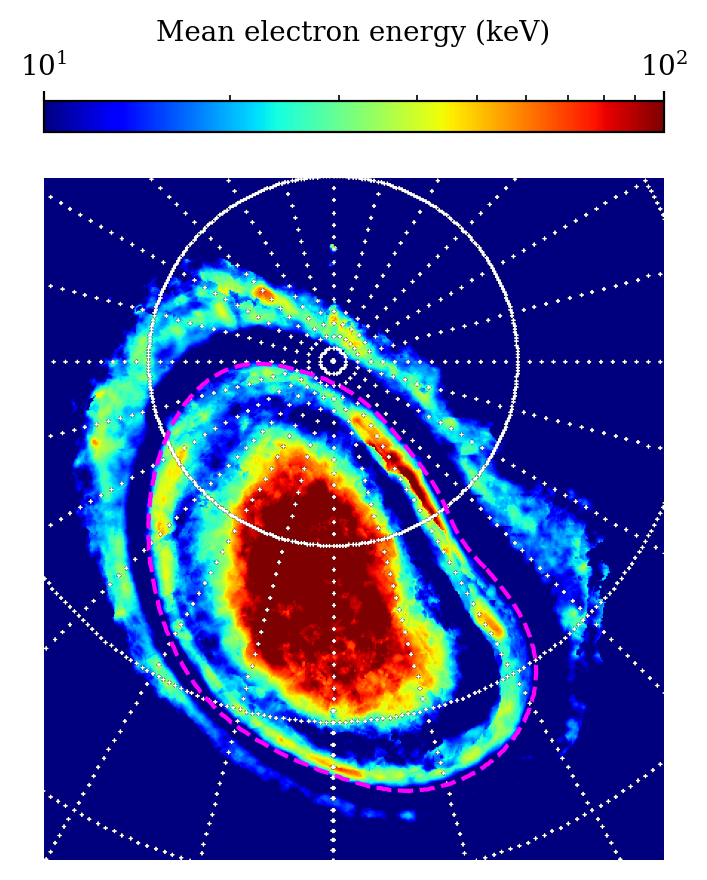}
    \caption{
        Stacked average map of mean auroral electron energy inferred from UVS in the northern hemisphere from PJ1 to PJ60. Annotation is as per Figure \ref{fig:map_uv_flux_N}.
    }
    \label{fig:map_uv_energy_N}
\end{figure}
\begin{figure}[bt]
    \centering
    \includegraphics[width=\linewidth]{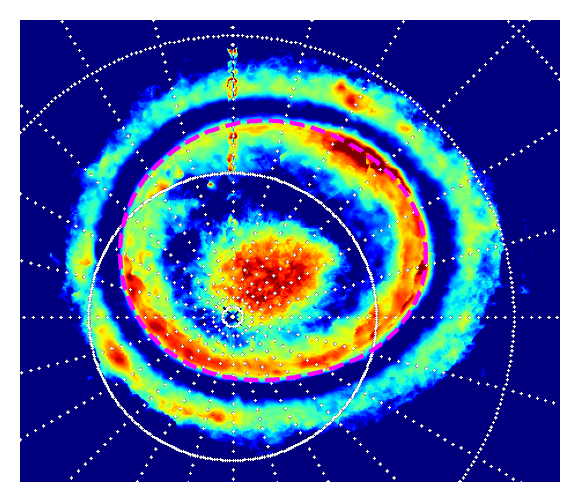}
    \caption{
        As Figure \ref{fig:map_uv_energy_N} but for the southern hemisphere.
    }
    \label{fig:map_uv_energy_S}
\end{figure}

Figures \ref{fig:map_uv_flux_N} to \ref{fig:map_uv_energy_S} show the log-mean (that is, the pixelwise mean of the log$_{10}$ of each map) UV-energy-flux and mean-electron-energy maps between PJ1 and PJ60 for the northern and southern hemispheres.
The main emission is clearly visible in both hemispheres and in both the energy-flux and mean-energy maps, alongside the diffuse aurora toward the equator \citep{radioti+:2009} and the polar-cap aurora \citep{mauk+:2020} that forms a distinct high-electron-energy region interior of the main emission.
The average main emission shows a noticeably greater energy flux in the south, whereas the polar-cap aurora is more intense in the north, in line with previous observations \citep{gerard+:2013,grodent+:2018,head+:2024}
The main-emission energy flux is reasonably uniform along its contour, especially in the southern hemisphere; this is as expected, since the known dawn-dusk asymmetry in main-emission brightness \citep{groulard+:2024,head+:2024} is removed when averaging the aurora in System-III longitude.
However, in the northern hemisphere, the brightest regions of the main emission appear to correspond with the region of higher surface magnetic field strength at a System-III longitude of $\sim$190\textdegree\ \citep{connerney+:2022}.
The main-emission energy-flux distribution is also well centred on the \citet{head+:2024} main-emission contour; this is unsurprising since this contour is based on the average main-emission brightness profile from PJ1 through PJ54.
More surprising, however, is the consistent and considerable poleward shift of the mean-energy profile in Figures \ref{fig:map_uv_energy_N} and \ref{fig:map_uv_energy_S} at all positions along the main-emission contour.

\begin{figure}[tb]
    \centering
    \includegraphics[width=\linewidth]{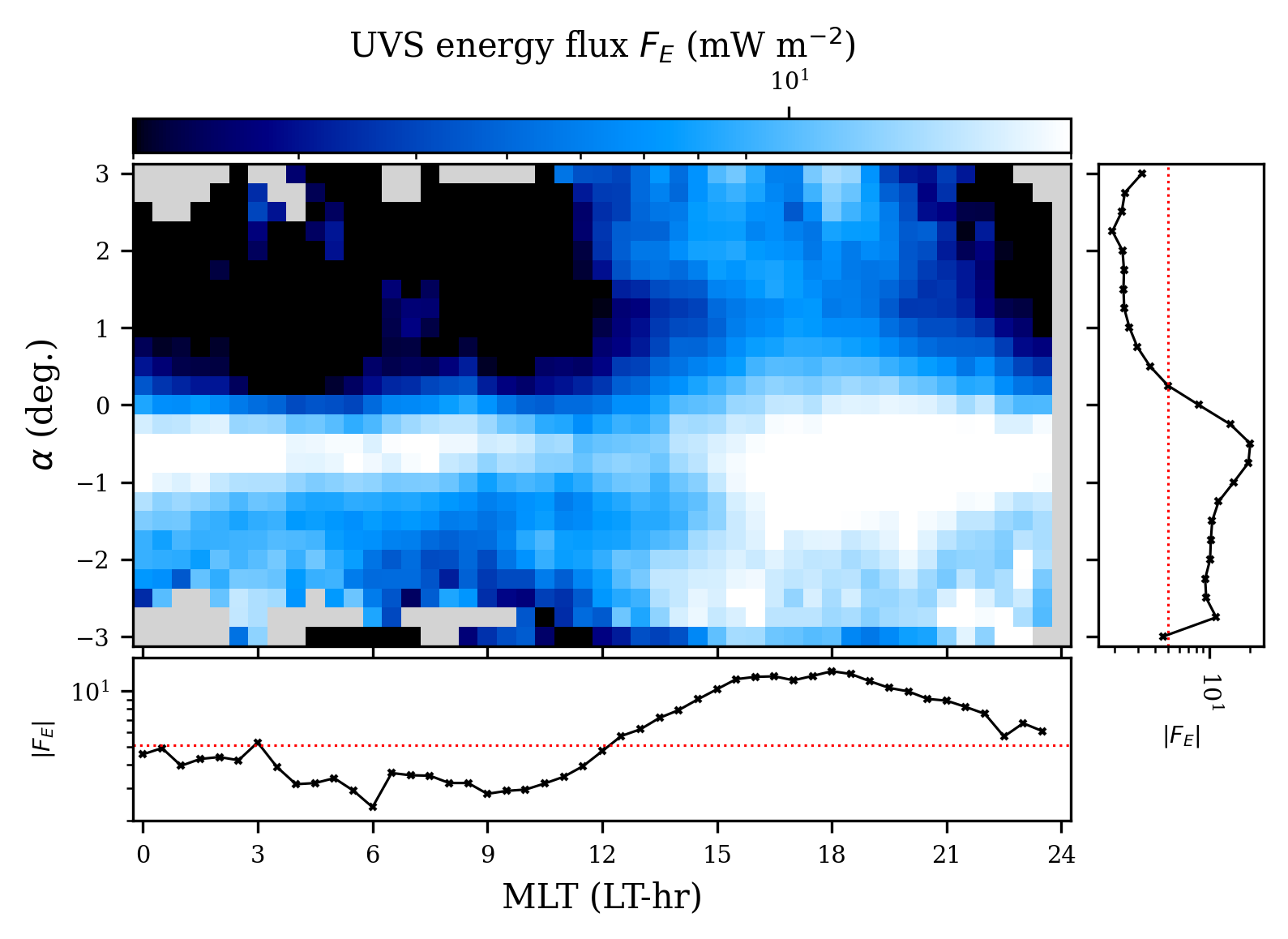}
    \caption{
        Stacked average map of auroral electron energy flux inferred from UVS in the main emission from PJ1 to PJ60, projected into the magnetosphere in magnetic local time and separation angle $\alpha$, where a positive $\alpha$ indicates a poleward deviation from the central profile of the main-emission mask. The (log-)mean latitudinal and local-time profiles are given to the right and below the average map respectively. The median value of these two profiles is denoted by the dotted red line. Grey pixels denote those regions without data.
    }
    \label{fig:map_uv_flux_proj}
\end{figure}
\begin{figure}[tb]
    \centering
    \includegraphics[width=\linewidth]{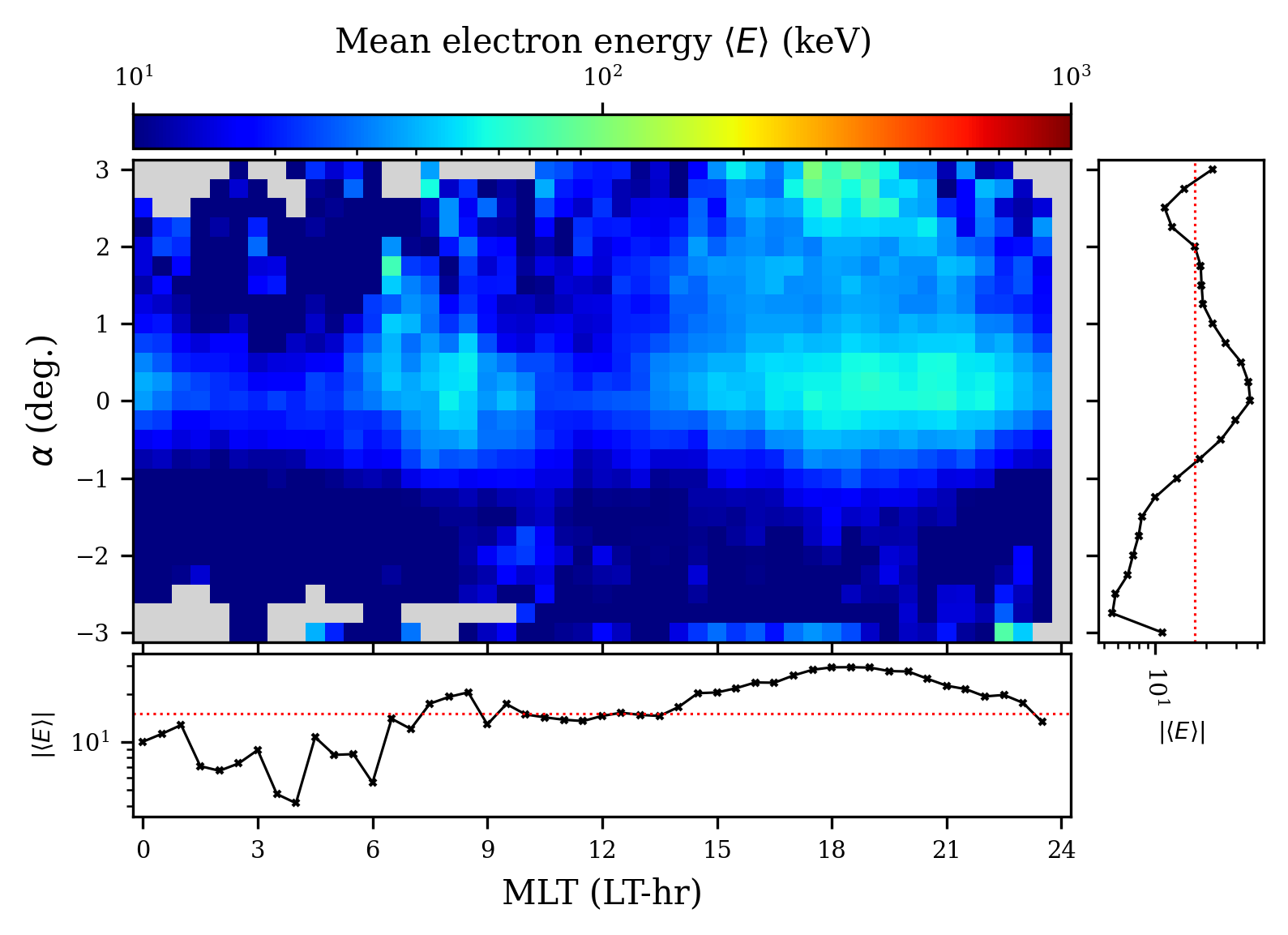}
    \caption{
        Stacked average map of mean precipitating electron energy inferred from UVS in the main emission from PJ1 to PJ60, projected into the magnetosphere in magnetic local time and separation angle $\alpha$. Further annotation is as per Figure \ref{fig:map_uv_flux_proj}.
    }
    \label{fig:map_uv_energy_proj}
\end{figure}

Figures \ref{fig:map_uv_flux_proj} and \ref{fig:map_uv_energy_proj} show the main-emission electron energy flux and mean electron energy for PJ1 to 60 in both hemispheres but instead binned by magnetic local time (the local time of the mapped location of the auroral pixel in the equatorial current sheet; variations in the surface magnetic-field strength are thus averaged out) and the deviation angle $\alpha$ (analogous, but not identical,  to the $\alpha$ parameter used in \citealt{kruegler+:2025}) from the local central axis of the main-emission mask, where this latter parameter has been determined by applying the \texttt{skimage.morphology.skeletonize} skeletonisation algorithm \citep{vanderwalt+:2014} to the main-emission mask.
The angle $\alpha$ has been used here instead of the mapped magnetospheric equatorial distance or deviation from a statistical main-emission contour to account for the variable size of the main emission \citep{head+:2024}.
It should be noted that the deviation angle $\alpha$ may also be defined based on in-situ measurements rather than the position of the UV main emission, though it was found that both definitions produce results that are in broad agreement (\citealt{kruegler+:2026}, Figure 3).
Here, unlike in Figures \ref{fig:map_uv_flux_N} and \ref{fig:map_uv_flux_S}, the local-time asymmetry of the main emission is clearly visible, insofar as the dusk-side main-emission is approximately twice as bright on average than the dawn-side main emission, in line with previous work \citep{head+:2024}.
At high $\alpha$ (poleward) in the dusk-side main emission, the presence of bridges, which are concentrated around 15-18 MLT \citep{head+:2025}, is clearly visible in both the energy-flux and mean-energy maps.
The reduction in average brightness around noon is also consistent with the known pre-noon discontinuity in the main emission \citep{radioti+:2008}.
The region of increased electron energy between 6 and 9 MLT in Figure \ref{fig:map_uv_energy_proj} is likely associated with (pseudo) dawn storms, despite the fact that the auroral images showing the most obvious dawn storms were removed from the dataset and not used to produce Figures \ref{fig:map_uv_flux_proj} and \ref{fig:map_uv_energy_proj}.
The latitudinal profile in Figure \ref{fig:map_uv_flux_proj} takes the form of an asymmetrical profile with a broad peak approximately in the centre of the main-emission mask. 
The shallower profile at negative $\alpha$ is likely due to the contribution of the diffuse aurora and injection signatures \citep{dumont+:2014} equatorward of the main emission.
In any case, the position of the peak around $\alpha=0^{\circ}$ is expected since the main-emission masks are produced by fitting a Gaussian profile to the energy-flux maps.
The slight equatorward shift ($\sim$0.5\textdegree) is likely due to the inclusion of bridges in the main-emission mask which would naturally bring the central axis of the mask poleward.
Here, like in Figures \ref{fig:map_uv_flux_N} to \ref{fig:map_uv_energy_S}, the mean-energy peak in Figure \ref{fig:map_uv_energy_proj} is consistently and considerably poleward of the energy-flux peak at all local times.
If only the main central profile of the main emission around $\alpha=0^{\circ}$ is considered, the mean-energy peak is nevertheless consistently $\sim$1\textdegree\ poleward of the energy-flux peak.
This is important, since it indicates that the more-poleward latitudinal profile in Figure \ref{fig:map_uv_energy_proj} is not simply due to the contribution of bridges in the dusk-side aurora beyond $\alpha=1^{\circ}$ but rather a ubiquitous characteristic of the main emission (although the inclusion of the bridges in the drawing of this conclusion may be justified, as they potentially form part of the Zone-II main emission; \citealt{head+:2025}).

This relationship between precipitating energy flux $F_E$ and mean electron energy $\langle E \rangle$ in the main emission has previously been modelled as a power law of the form 
\begin{equation}
    \label{eq:power_law}
    \langle E \rangle=10^{A}\cdot F_E^k,
\end{equation}
where $k$ and $A$ are factors that are given by the gradient and intercept of a fitted linear relation to $F_E$ and $\langle E \rangle$ in log space \citep{clark+:2018}.
The choice of $F_E$ or $\langle E \rangle$ as the independent variable in this power law is somewhat arbitrary; work focused on in-situ data tends to use $\langle E \rangle$ as the independent variable \citep{clark+:2018} which is perhaps more physically meaningful, whereas previous remote-sensing work uses $F_E$ as the independent variable \citep{gustin+:2004,gerard+:2016} since energy flux is the quantity that is more directly accessible from imaging data, such as those considered in this work.
It is notable that both the quasi-static-potential \citep[e.g.][]{cowley+bunce:2001} and Alfvénic-turbulence \citep{saur+:2018} mechanisms lead to (approximate) power-law relations that are compatible with Juno observations of precipitating electrons above the main emission \citep{clark+:2018}.
In the case of Alfvénic turbulence, the expected power law is 
\begin{equation}
    \label{eq:power_law_alfvenic}
    \langle E \rangle \propto n_e^{[-0.75,-0.5]}F_E^{0.5},
\end{equation}
where $n_e$ is the local electron number density in the ionosphere.
The dependence on density $n_e$ is expected to have a range of exponents $[-0.75, -0.5]$ based on the degree of relativisticness of the system; see Appendix \ref{app:power_law_derivation}.
In the case of acceleration by a quasi-static electric potential, the power law is instead 
\begin{equation}
    \label{eq:power_law_currents}
    \langle E \rangle \propto n_e^{-0.5}F_E^{0.5}.
\end{equation}
Details of the derivations of these two relations, as well as any simplifying assumptions, are given in Appendix \ref{app:power_law_derivation} and \citet{clark+:2018}.
$k$ is thus expected to be the same for the two relations, whereas $A$ is expected to have a slightly different dependence on the ionospheric electron number density $n_e$.
In both cases, as the ionospheric electron density decreases, the average electron energy for a given electron flux is expected to increase. 

\begin{figure}[tb]
    \centering
    \includegraphics[width=\linewidth]{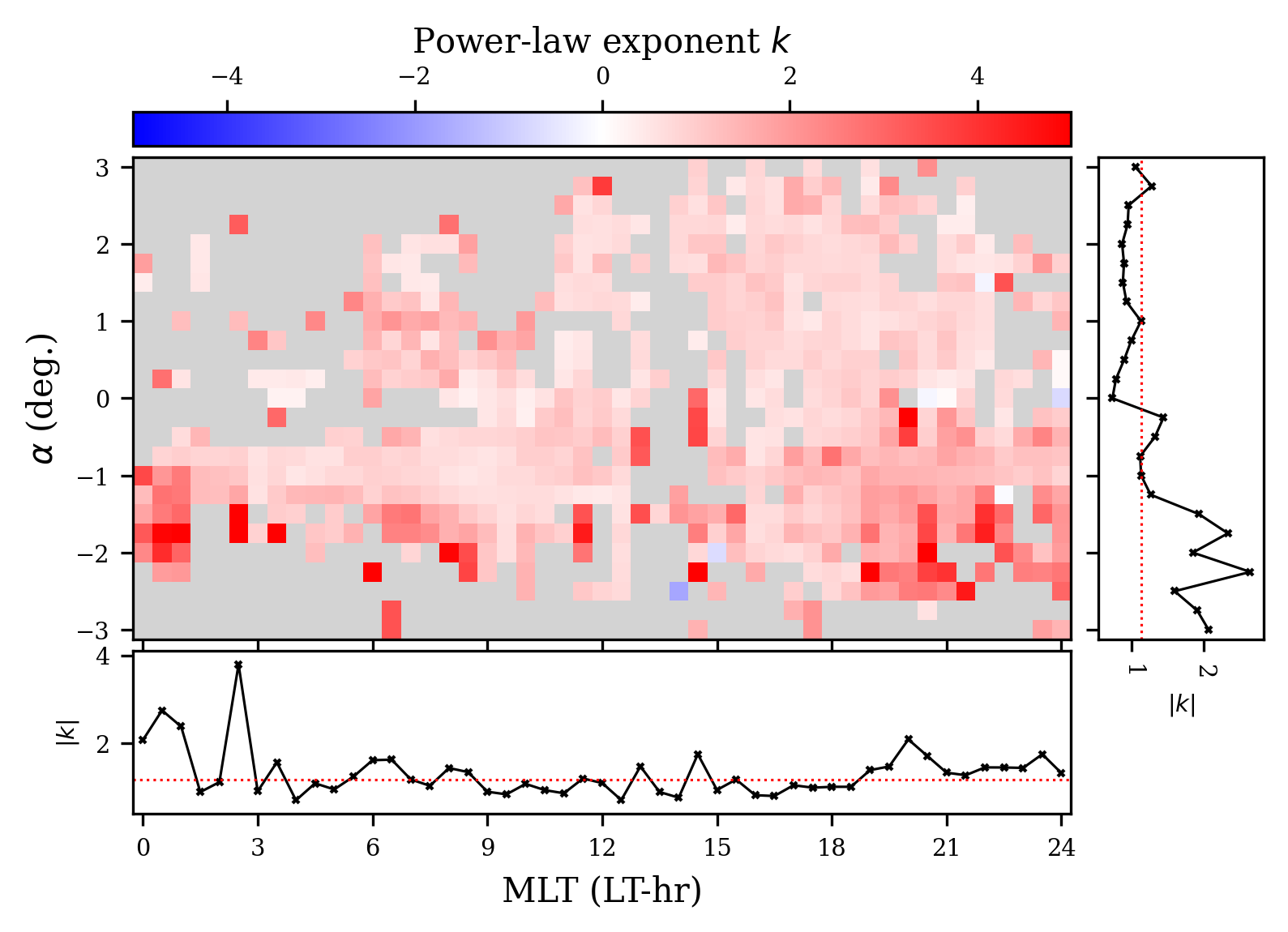}
    \caption{
        Stacked average map of the fitted power-law exponent $k$ from Equation \ref{eq:power_law} in the main emission from PJ1 to PJ60, projected into the magnetosphere in magnetic local time and separation angle $\alpha$. Further annotation is as per Figure \ref{fig:map_uv_flux_proj}. 
    }
    \label{fig:map_k}
\end{figure}
\begin{figure}[tb]
    \centering
    \includegraphics[width=\linewidth]{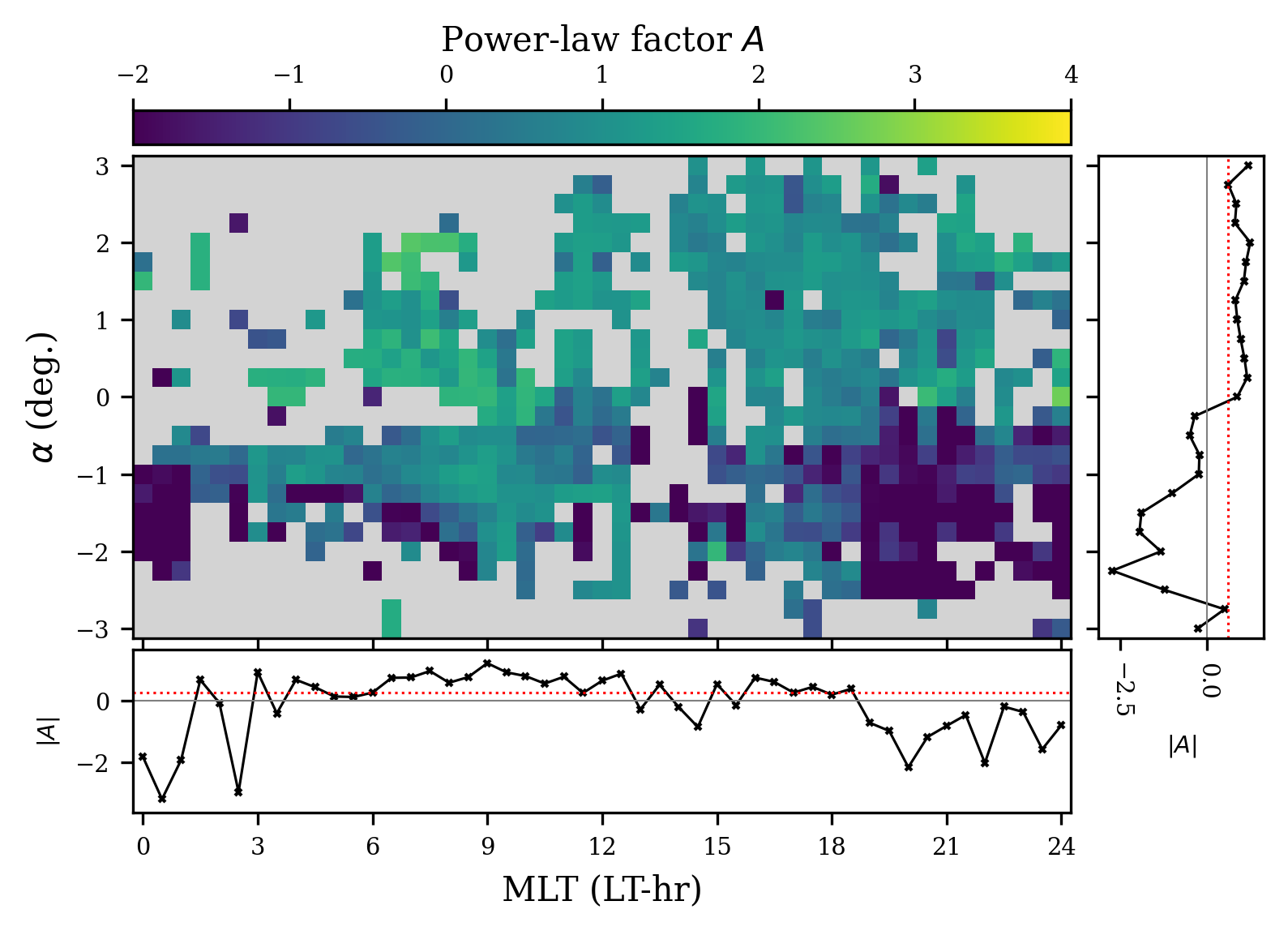}
    \caption{
        Stacked average map of the fitted power-law factor $A$ from Equation \ref{eq:power_law} in the main emission from PJ1 to PJ60, projected into the magnetosphere in magnetic local time and separation angle $\alpha$. Further annotation is as per Figure \ref{fig:map_uv_flux_proj}. 
    }
    \label{fig:map_A}
\end{figure}

Figures \ref{fig:map_k} and \ref{fig:map_A} show the distribution of power-law factors $k$ and $A$ within the main emission.
Here, each image of the aurora was separated into the full number of bins in magnetic local time and $\alpha$, and the power-law factors $k$ and $A$ fitted to the pixels contained within this bin and its four neighbouring bins, to ensure sufficient numbers of pixels to fit a representative power law.
Only power-law fits with R$^{2}$ values above 0.7 are included to ensure that only reliable power laws within the aurora are analysed.
Figures \ref{fig:map_k} and \ref{fig:map_A} contain the bin-wise averages, over all the auroral images considered in this work, of $k$ and $A$ respectively.
Figure \ref{fig:map_k} shows an elevated $k$ in the region of $\alpha<-1^{\circ}$ followed by a plateau at higher $\alpha$ at a value of 0.81$\pm$0.13 which is broadly consistent with the expected power-law exponent of $k=0.5$ for a main emission dominated by acceleration from either quasi-static potentials or Alfvénic turbulence \citep{clark+:2018} and compatible with previous results based on HST measurements \citep{gerard+:2016}.
Figure \ref{fig:map_A} shows a similar drop in $A$ in the region of $\alpha<-1^{\circ}$ which is instead followed by a region of increased $A$ with increasing $\alpha$.
Both figures show relatively flat local-time profiles, indicating that both $k$ and $A$ are comparable at all local times in the main emission.
There appears to be a decrease in $A$ in the dusk-side main emission around $\alpha=-1^{\circ}$; since the average value of $A$ in the bridge region (dusk, $\alpha>0^{\circ}$) is similar to the value in the dawn-side main emission, we interpret this as a consequence of the much clearer separation of the Zone-I and Zone-II main emission in the dusk sector.
We interpret the sharp gradients in the region of $\alpha<-1^{\circ}$ to be due to the influence of injection signatures and/or the diffuse aurora, which can occur on or just equatorward of the main emission.
Thus, we suggest that our observations are broadly compatible with a main-emission flux-energy power law of exponent $k\sim0.5$ at all (magnetic) latitudes and local times, whereas the factor $A$ increases with increasing latitude.

\subsection{Analysis of in-situ data}
\label{sec:results_insitu}

In addition to the results available via remote sensing of Jupiter's aurora detailed in section \ref{sec:results_remote}, we can also combine UVS data with data from Juno's in-situ instruments (MAG, JEDI) to further investigate the average properties of the main emission.
It should first be noted that the characteristic (that is, the log-average of the log-average of each parameter observed at Juno's footprint during each crossing) energy fluxes and mean electron energies for the two zones were found from UVS data to be 17$_{-12}^{+42}$ mW m$^{-2}$ and 43$_{-29}^{+94}$ keV for Zone-I and 8$_{-7}^{+44}$ mW m$^{-2}$ and 199$_{-168}^{+1086}$ keV for Zone-II.
We identify that the Zone-I and Zone-II have comparable energy fluxes (and hence that both zones meaningfully contribute toward the main emission) but that the poleward Zone-II has a noticeably elevated mean electron energy, as suggested by Figure \ref{fig:map_uv_energy_proj}.
Although the Zone-I (upward-current) aurora is most frequently found equatorward of the Zone-II (downward-current) aurora \citep{mauk+:2020}, several cases have previously been identified in which the two zones were instead inverted in latitude, though only in the northern aurora, which was taken to be evidence of temporary super-corotation of the middle magnetosphere and hence an inversion of the typical current loop \citep{al_saati+:2022}. 
However, the conjugate scenario (latitudinally inverted Zone-I/II in the southern aurora) was not identified by \citet{al_saati+:2022}.
This work proposes two cases (PJ14-S and PJ18-S) in which the Zone-I/II were indeed inverted in the south, which are presented in Appendix \ref{app:z12_south_inversion}.
This result completes the set of observed Zone-I/II configurations in both hemispheres and indicates that the conclusions drawn by \citet{al_saati+:2022} based solely on the behaviour of the northern aurora do seem to be more widely applicable to the full aurora/magnetosphere system.
Interestingly, PJ14-N was one of the inverted cases identified by \citet{al_saati+:2022}, whereas PJ18-N does appear to retain the typical ``Zone-II poleward'' configuration, indicating that the super-corotation (or whatever other process) that leads to the inversion of these two zones may, but is not required to, appear or disappear over the hour-long timescale between Juno crossings of the northern and southern aurora.
It cannot be excluded that further cases of this inversion exist, especially in later perijoves where the identification of the Zone-I and Zone-II aurora is more difficult, but PJ14-S and PJ18-S represent the most obvious and credible cases.
PJ14 and PJ18 were indeed analysed by \citet{al_saati+:2022}, but it is expected that they were not previously identified as inverted cases due to the presence of features in the field-aligned-current data (see Appendix \ref{app:z12_south_inversion}) that make proper identification of the Zone-I/II aurora difficult without verification against the instantaneous morphology of the UV aurora and the mapped location of the Juno magnetic footprint.  
The full list of cases displaying this zonal inversion is given in Appendix \ref{app:crossing_table}; there appears to be no correlation with the state of compression of the magnetosphere, as inversions are noted to have occurred when the main emission was both very expanded (PJ22-N) and very contracted (PJ10-N) compared to its average position \citep{head+:2024}.
This is not in line with the expected behaviour of a current-based main emission, in which solar-wind compression is expected to lead to super-corotation on closed field lines and an inversion of the global current-loop system \citep{cowley+:2007}.
Alternatively, instead of an inversion of the global current-loop system, these cases may instead represent the crossing of a more-complex ``folded'' current-loop morphology, as has been observed on Earth \citep{ivarsen+:2024}, or may simply be a misinterpretation based on the approximate Zone-I/II identification conditions noted in section \ref{sec:observations} or oversimplification of the current-density calculation \citep{head+:2025}.

In this work, we identify 10 discrete inverted-V structures in the downward-electron energy spectra observed by JEDI during Juno crossings of the main emission.
A full list of these cases is given in Appendix \ref{app:crossing_table}.
These represent 11\% of the 92 main-emission crossings (118 hemisphere traversals from PJ1-60, excluding cases with insufficient UVS coverage or no main-emission crossing) considered in this work, which is comparable to the relative occurrence rate of 7\% identified above the main emission by \citet{salveter+:2022} relative to the more-common broadband distributions, though \citet{salveter+:2022} suggest that this occurrence rate be an underestimate of the true occurrence rate of inverted-V structures above the aurora as they only considered electron spectra (and hence could only detect inverted-Vs above Juno).
Inverted-Vs below Juno would be detectable in the upward-travelling proton spectra which were not considered by \citet{salveter+:2022} nor in this work. 
Interestingly, in cases where the Zone-I/II aurora could be reliably identified, these inverted-V structures were seen exclusively above the Zone-I aurora, as indicated by previous work based on a more limited number of perijoves \citep{mauk+:2020,sulaiman+:2022}, though this result is biased toward the dusk- and night-side aurora and is not necessarily representative of the full main emission.
Above the Zone-II aurora, no inverted-V structures were seen either in the downward or upward JEDI electron spectra, even in cases where an inverted-V was seen above the Zone-I aurora.
Of the cases noted in Table \ref{tab:crossing_table}, all but one (PJ14-N) presented a main emission that was contracted compared to its average position, indicative of compression of the magnetosphere by the solar wind \citep{head+:2024} and hence also that this compression may play a role in the generation of these inverted-V potential structures seen occasionally above the main emission, though it must be noted that there are a large number of cases with a contracted main emission that did not display an inverted-V structure.


\begin{figure}[tb]
    \centering
    \includegraphics[width=\linewidth]{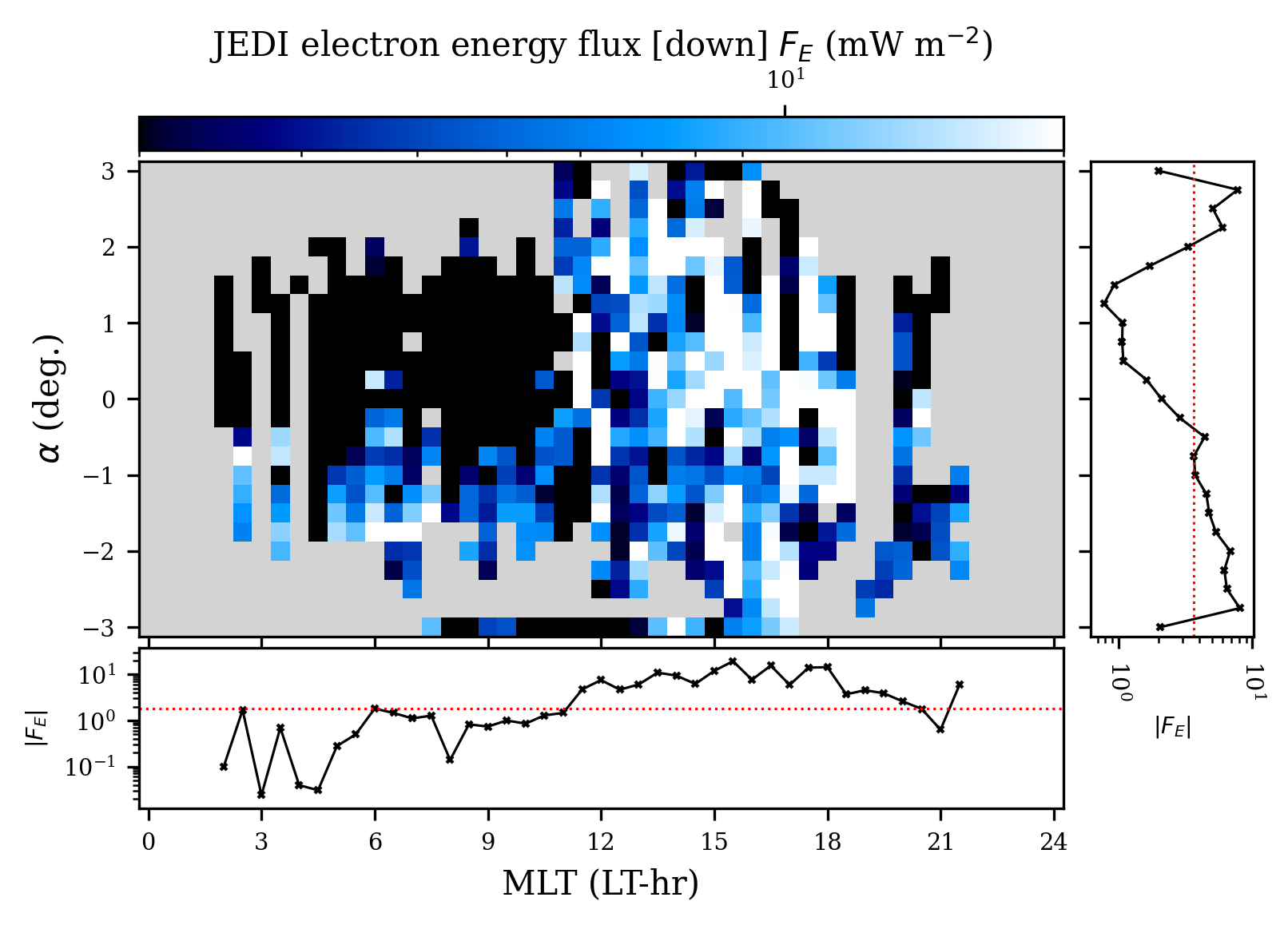}
    \caption{
        Stacked average map of downward electron energy flux measured by JEDI above the main emission from PJ1 to PJ60, projected into the magnetosphere in magnetic local time and separation angle $\alpha$. Further annotation is as per Figure \ref{fig:map_uv_flux_proj}.
    }
    \label{fig:map_jedi_flux_proj}
\end{figure}

\begin{figure}[tb]
    \centering
    \includegraphics[width=\linewidth]{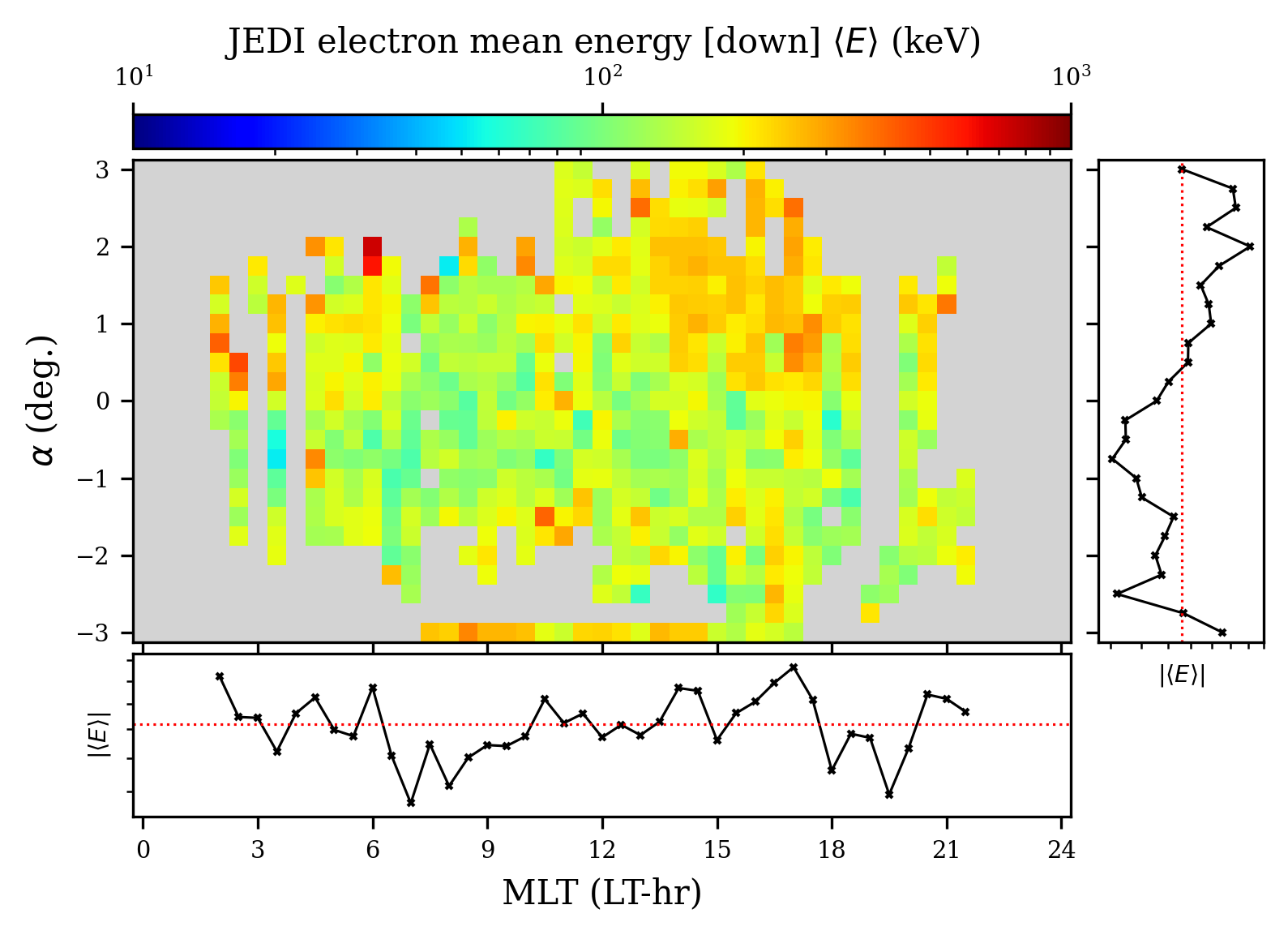}
    \caption{
        Stacked average map of downward electron mean energy measured by JEDI above the main emission from PJ1 to PJ60, projected into the magnetosphere in magnetic local time and separation angle $\alpha$. Further annotation is as per Figure \ref{fig:map_uv_flux_proj}.
    }
    \label{fig:map_jedi_energy_proj}
\end{figure}

Figures \ref{fig:map_jedi_flux_proj} and \ref{fig:map_jedi_energy_proj} show the energy flux and mean energy of downward-travelling electrons for Juno crossings of the main emission below 2 R$_{J}$.
Here, the maximum electron pitch angle has been fixed at 20\textdegree\ of the local magnetic-field vector, which is the expected loss-cone angle at an altitude of 1 R$_{J}$ and thus broadly consistent with the crossings presented in this work, which have a minimum altitude of 0.05 R$_{J}$ and have been restricted to those crossings below 2 R$_{J}$.
In any case, since acceleration from either quasi-static potentials or Alfvén waves is expected to both be field-aligned and to dominate the electron energy flux within the loss cone, it is required that the choice of maximum pitch angle contains the full field-aligned electron flux but not necessarily exclusively.
The validity of this assumption is discussed further in Appendix \ref{app:pitch_angle_check}.
We can see that Figures \ref{fig:map_jedi_flux_proj} and \ref{fig:map_jedi_energy_proj} show approximately similar trends as Figures \ref{fig:map_uv_flux_proj} and \ref{fig:map_uv_energy_proj}: the energy flux shows a profile dominated by the contribution below $\alpha=0^{\circ}$ whereas the mean energy shows a distinct increase at greater $\alpha$.
Differences between the latitudinal profile in Figure \ref{fig:map_jedi_flux_proj} (Figure \ref{fig:map_jedi_energy_proj}) and Figure \ref{fig:map_uv_flux_proj} (Figure \ref{fig:map_uv_energy_proj}) are likely strongly influenced by the comparative lack of coverage in the in-situ JEDI data.
The average downward electron energy flux measured by JEDI is of similar magnitude to that inferred from UVS ($\sim$10 mW m$^{-2}$), whereas the average mean electron energy measured by JEDI is full order of magnitude above the average extrapolated from UVS observations.
This is expected to be due to the lack of JADE data used in this work; the average energy measured by JEDI alone will naturally always be within the 20 - 1000 keV range, whereas, in reality, there may be a significant contribution from <20 keV electrons that would be captured in the UVS mean-energy maps \citep{ebert+:2019,benmahi+:2026}.
The quantitative agreement the UVS and JEDI energy fluxes is discussed further in Appendix \ref{app:uvs_jedi_flux}, where it is suggested that the reduction in JEDI energy flux observed at higher altitudes is consistent with the existence of an auroral acceleration region at an altitude of $\lesssim$1 R$_{J}$ \citep{hess+:2010,sulaiman+:2022}. 
With this in mind, we would therefore expect the latitudinal profiles in Figures \ref{fig:map_uv_energy_proj} and \ref{fig:map_jedi_energy_proj} to have similar shapes but not necessarily the same average values, depending on the relative contribution of 20 - 1000 keV electrons for a given location in the aurora. 
The dawn-dusk asymmetry present in the energy flux in Figure \ref{fig:map_uv_flux_proj} is also present in Figure \ref{fig:map_jedi_flux_proj}.
In all, these results indicate that the energy fluxes and mean electron energies extracted from UVS data are broadly compatible with the properties of downward-travelling electrons measured by JEDI, and thus that these latter electrons are indeed precipitating and producing the main emission.


\begin{figure}[tb]
    \centering
    \includegraphics[width=\linewidth]{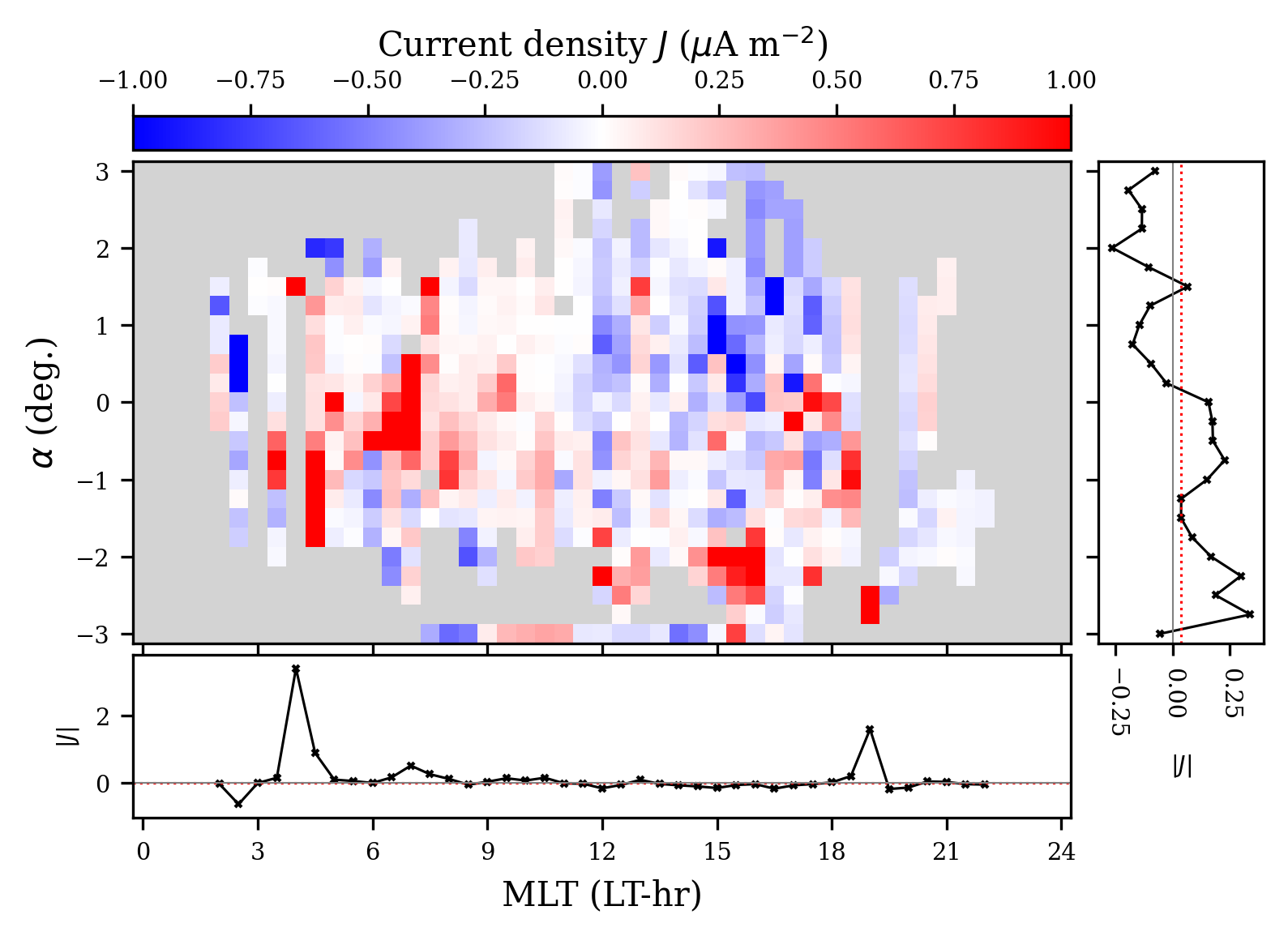}
    \caption{
        Stacked average map of ionospheric field-aligned currents extrapolated from Juno-MAG above the main emission from PJ1 to PJ60, projected into the magnetosphere in magnetic local time and separation angle $\alpha$. Further annotation is as per Figure \ref{fig:map_uv_flux_proj}. Positive values (red) denote upward field-aligned currents.
    }
    \label{fig:map_currents}
\end{figure}

It may be tempting to associate the higher-$\alpha$ mean-energy peak shown in Figures \ref{fig:map_uv_energy_proj} and \ref{fig:map_jedi_energy_proj} with the Zone-II of the main emission, which is usually found in the poleward part of the full main-emission profile \citep{al_saati+:2022}.
To make this association more robust, we can consider the average field-aligned current density seen above the main emission, extrapolated to the expected auroral altitude of 400 km \citep{head+:2025}, during Juno crossings, which are given in Figure \ref{fig:map_currents}.
Here, the average latitudinal profile shows a consistently upward current at $\alpha<0^{\circ}$ and a consistently downward current at $\alpha>0^{\circ}$, consistent with the Zone-I and Zone-II main emission respectively \citep{mauk+:2020,kruegler+:2026}.
Since this is an average profile, it cannot be used to say whether Juno is above the Zone-I or Zone-II main emission during a given main-emission crossing; however, it can be said that there is an increased probability of Juno being above the Zone-II aurora at greater $\alpha$ and especially when $\alpha>0^{\circ}$.

\begin{figure}[tb]
    \centering
    \includegraphics[width=\linewidth]{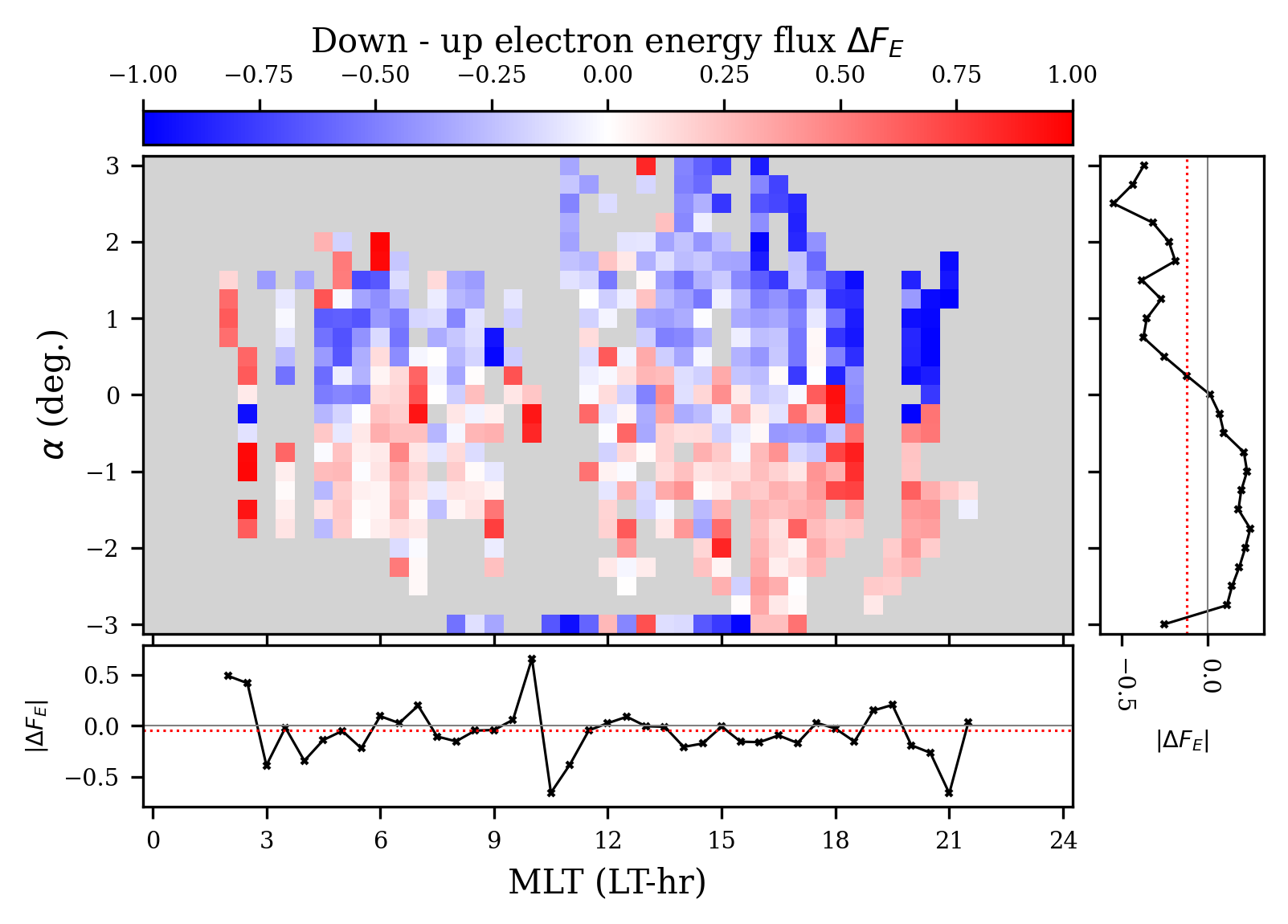}
    \caption{
        Stacked average map of the difference between the downward and upward electron energy flux seen by JEDI above the main emission from PJ1 to PJ60, projected into the magnetosphere in magnetic local time and separation angle $\alpha$. Further annotation is as per Figure \ref{fig:map_uv_flux_proj}. Positive values (red) denote regions dominated by downward electrons.
    }
    \label{fig:map_fluxdiff}
\end{figure}

The (normalised) difference between the downward and upward loss-cone electron energy flux, shown in Figure \ref{fig:map_fluxdiff}, allows us to draw a similar conclusion.
Here, as in Figure \ref{fig:map_currents}, the main emission equatorward of $\alpha=0^{\circ}$ is dominated by downward electrons and the main emission poleward of $\alpha=0^{\circ}$ instead by upward electrons, as would be expected from the Zone-I and Zone-II aurora respectively.
We suggest that, for a given crossing of the main emission, Juno is more likely to be above the Zone-I aurora when $\alpha<0^{\circ}$ and above the Zone-II aurora when $\alpha>0^{\circ}$.
With a sufficiently large dataset, such as the 60 perijoves discussed in this work, this allows us to draw statistical conclusions about the Zone-I and Zone-II aurora based solely on the mapping of Juno within the main emission, even in the absence of in-situ signatures of these zones, which, as discussed above, are frequently absent especially toward later perijoves.
Naturally, the identification of the Zone-I and Zone-II aurora from observational signatures, rather than this statistical argument, should be prioritised when possible.
Figures \ref{fig:map_currents} and \ref{fig:map_fluxdiff} provide evidence that the consistently poleward peak of the mean precipitating electron energy (Figures \ref{fig:map_uv_energy_proj} and \ref{fig:map_jedi_energy_proj}) compared to the energy-flux peak (Figures \ref{fig:map_uv_flux_proj} and \ref{fig:map_jedi_flux_proj}) is indeed associated with the Zone-II aurora, rather than the Zone-I aurora or a combination of the two.
Conversely, the brightness peak at $-1^{\circ} < \alpha < 0^{\circ}$ in Figure \ref{fig:map_uv_flux_proj} may indicate that the Zone-I aurora is typically the brightest portion of the main emission.
Figures \ref{fig:map_currents} and \ref{fig:map_fluxdiff} also indicate that the power-law exponent $k$ is broadly uniform for the main emission beyond $\alpha>-1^{\circ}$ (where injection signatures are not expected to occur) and thus likely the same for the Zone-I and Zone-II aurora.

\begin{figure}[tb]
    \centering
    \includegraphics[width=\linewidth]{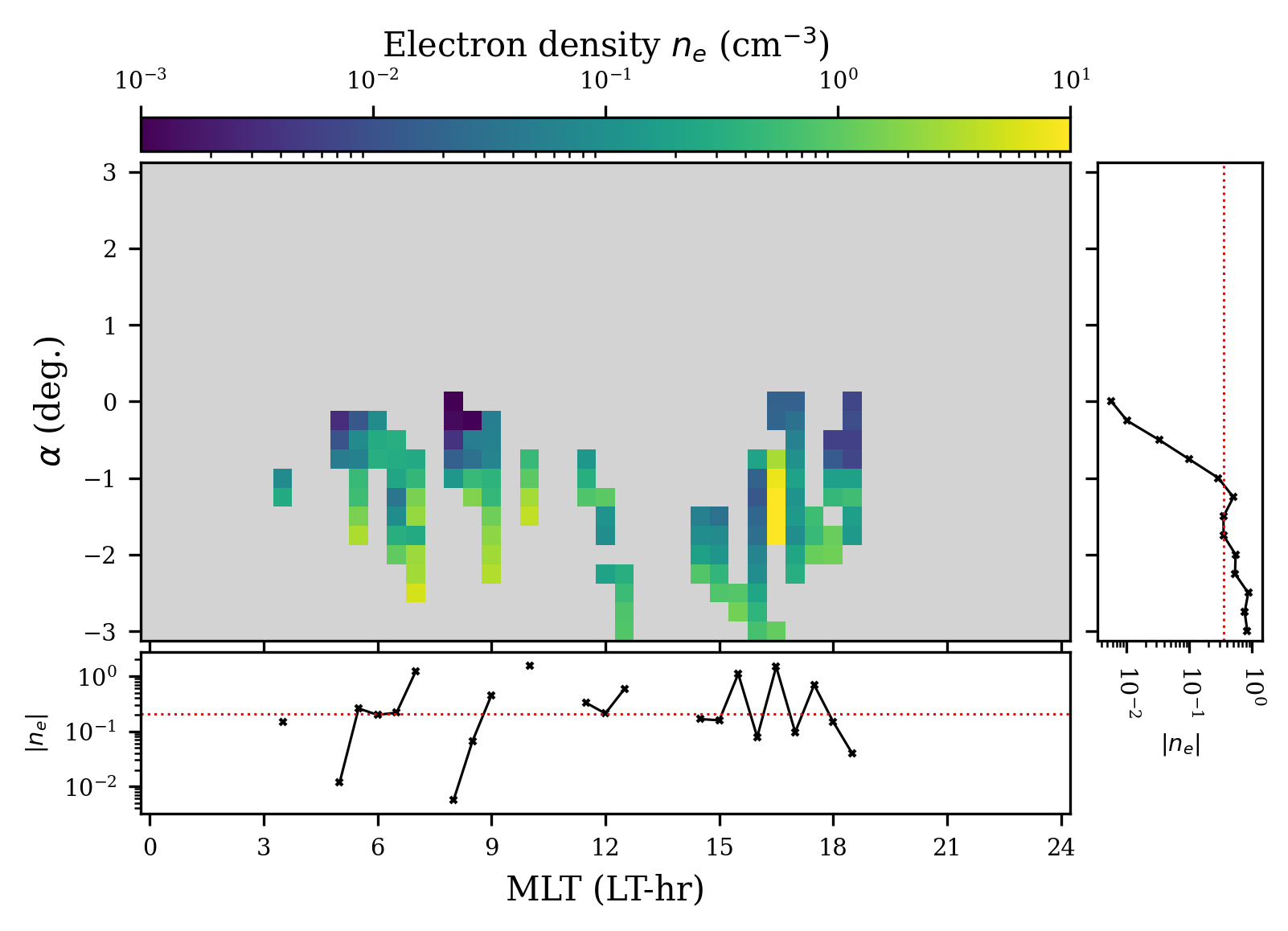}
    \caption{
        Stacked average map of the derived ionospheric electron number density $n_e$ from Juno crossings of the main emission from PJ1 to PJ60 \citep{kruegler+:2025}, projected into the magnetosphere in magnetic local time and separation angle $\alpha$. Further annotation is as per Figure \ref{fig:map_uv_flux_proj}. 
    }
    \label{fig:map_density}
\end{figure}

The power-law factor $A$ is expected to depend on the density of electrons in the ionosphere $n_e$, which can be inferred from frequency cutoffs present in the plasma-wave spectra observed during Juno crossings of the main emission \citep{kruegler+:2025}, as $A\propto n_e^p$, where $p$ depends on the dominant acceleration mechanism (see Appendix \ref{app:power_law_derivation}).
These derived density profiles (from \citealt{kruegler+:2025}) are plotted for each crossing of the main emission in Figure \ref{fig:map_density}. 
As expected from \citet{kruegler+:2025, kruegler+:2026}, the average density profile shows a steady decrease as Juno moves poleward over the main emission, which was interpreted as the erosion of the ionosphere from inertial Alfvén waves.
Reliable estimation of the ionospheric electron density becomes challenging when the ionospheric density is too low; the density profiles in Figure \ref{fig:map_density} are thus restricted to the (higher-density) equatorward ($\alpha<0^{\circ}$) main emission and hence likely the Zone-I aurora, corroborated by the results of \citet{kruegler+:2026}.
This allows us to say that the densities in the Zone-II aurora are likely reduced compared to the Zone-I aurora without being able to derive any reliable quantitative values.

\begin{figure}[tb]
    \centering
    \includegraphics[width=\linewidth]{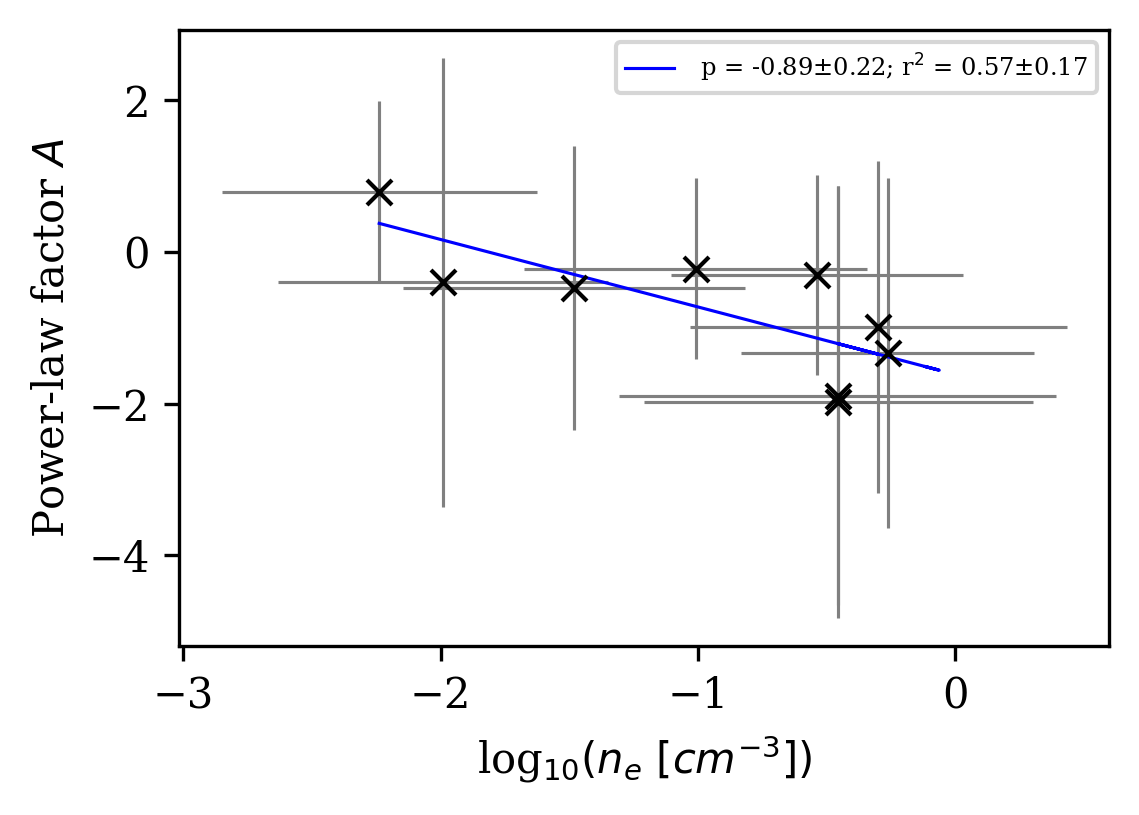}
    \caption{
        Average ionospheric electron density $n_e$ against average power-law factor $A$ for the latitudinal profiles presented in Figures \ref{fig:map_density} and \ref{fig:map_A} respectively. The fitted linear relation is given by a solid blue line. 
    }
    \label{fig:A_vs_density}
\end{figure}

Visually, the average latitudinal profile of the power-law factor $A$ in Figure \ref{fig:map_A} appears to be inversely proportional to the average density latitudinal profile in Figure \ref{fig:map_density}.
These two latitudinal profiles are plotted against one another in Figure \ref{fig:A_vs_density}.
It can be seen that a broadly inverse relation exists between the two parameters, as expected from Equations \ref{eq:power_law_alfvenic} and \ref{eq:power_law_currents}; indeed, the fitted exponent $p=-0.89\pm0.22$ matches well with the expected exponent for Alfvénic acceleration ($p=-0.75$).
However, given the relatively large scatter in the points (despite a reasonable R$^{2}$), we would caution against the use of Figure \ref{fig:A_vs_density} to support one acceleration mechanism over another.
Instead, since average profiles are used (instead of evaluating $A$ and $n_e$ on a case-by-case basis), we may say that, statistically, there exists an inverse relation between ionospheric electron density and the flux-energy power-law magnitude $A$ that is broadly (to an order of magnitude) compatible with previous theoretical predictions for either a current-based or turbulence-based main emission \citep{clark+:2018}.
A case-by-case analysis is included in Appendix \ref{app:density_cases} which shows a much weaker (though still negative) dependency between these two parameters.
We interpret this as an indicator that, while ionospheric electron density modifies the auroral flux-energy relation in a broad sense, other transient factors may dominate over the density dependence on a case-by-case basis.
It should also be stated that, due to the lack of reliable density estimates when $\alpha>0^{\circ}$, this relation can only be reliably said to hold for the Zone-I aurora.
\begin{figure}[tb]
    \centering
    \includegraphics[width=\linewidth]{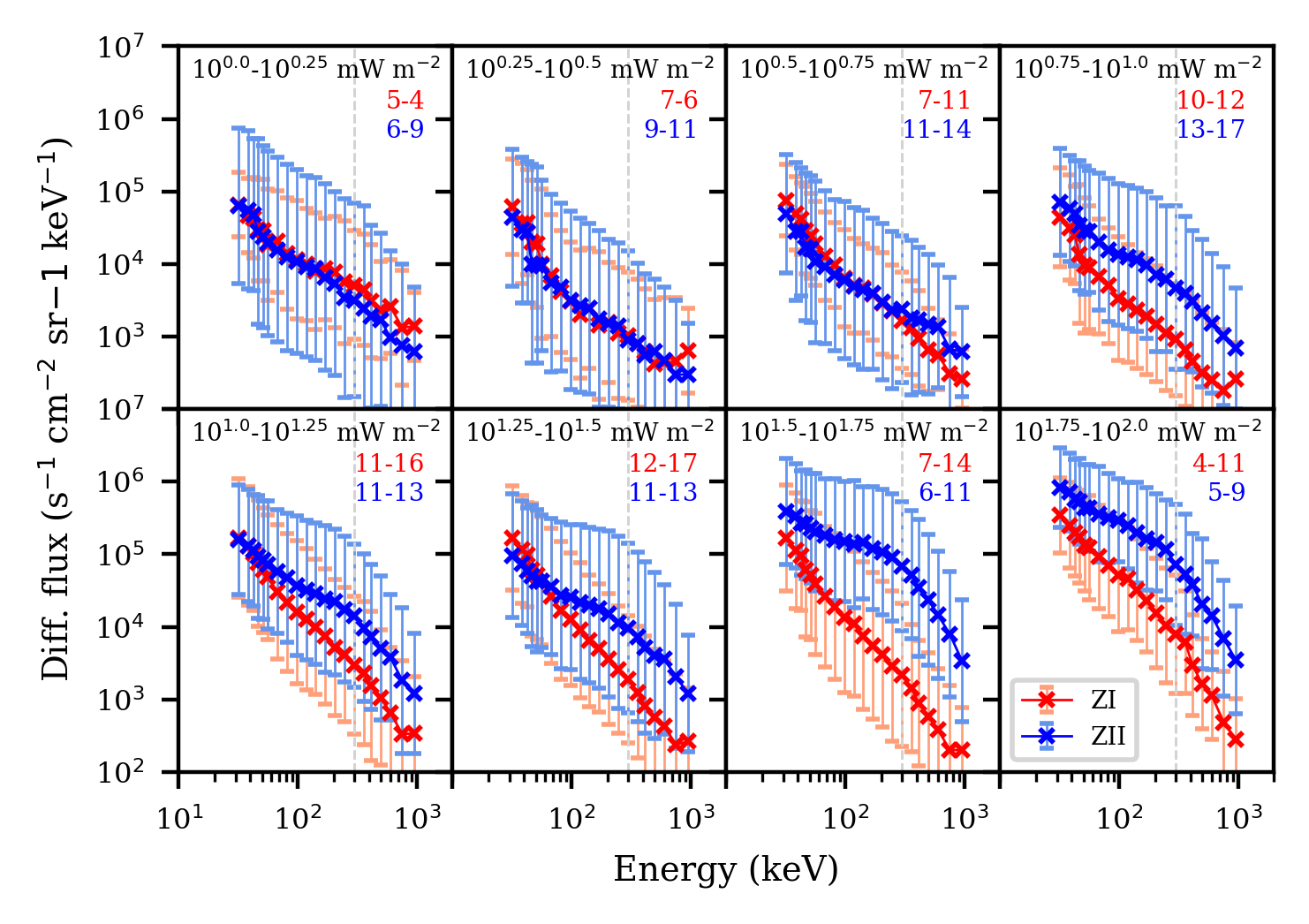}
    \caption{
        Average energy spectra (energy vs. differential electron flux) for Juno crossings of the Zone-I (red) and Zone-II (blue) main emission, binned by increasing (left to right) auroral electron energy flux. Errorbars represent the (log) standard deviation of values over all analysed crossings within each energy bin. A grey dashed line is included at 300 keV. The number of perijoves included in each bin is given in the format ``North-South'' in red for Zone-I then blue for Zone-II.
    }
    \label{fig:ebert2019_analogue}
\end{figure}

The energy spectra of precipitating electrons in the Zone-I/II main emission can be investigated directly using JEDI observations; a number of average profiles for the Zone-I/II aurora, binned by instantaneous electron energy flux inferred from UVS at the Juno footprint, are given in Figure \ref{fig:ebert2019_analogue}.
This is analogous to Figure 4 of \citet{ebert+:2019}, though they used JADE data (which detects electrons at lower energies than JEDI) in their analysis.
\citet{ebert+:2019} also only considered data when Juno was just poleward of the main emission over a region containing a bridge-like (Zone-II) arc, whereas Figure \ref{fig:ebert2019_analogue} presents the analysis separately for the Zone-I and Zone-II aurora.
We can see that, on average, at low energy fluxes, both the Zone-I and Zone-II aurora are associated with similar power-law-like distributions (straight lines in log space).
At higher energy fluxes, particularly beyond 10$^{1.25}$ mW m$^{-2}$, the average Zone-II spectrum presents a distinct inflection point or ``knee'' around 300 keV (grey dashed line in Figure \ref{fig:ebert2019_analogue}), which are referred to as ``plateau'' or ``flattop'' distributions \citep{clark+:2018}.
The Zone-I energy spectra remain clearly power-law-like even at high energy fluxes.
This difference in behaviour of the Zone-I and Zone-II precipitating-electron energy spectra is potentially indicative of two different acceleration mechanisms that give rise to the two zones of the main emission. 
This supports the idea that the Zone-II aurora significantly contributes towards the total emitted power of the main emission, as evidenced in previous work \citep{sulaiman+:2022, head+:2025}, especially from higher-energy electrons.
This is notable as the Zone-II aurora is dominated by upward-travelling electrons, which reinforces the idea that there must be significant bidirectional acceleration, perhaps via stochastic Alfvénic acceleration \citep{saur+:2018, sulaiman+:2022}, to produce the meaningful downward electron energy flux required to sustain this substantial contribution to the total main emission. 
The increased mean electron energy implied by the flattop Zone-II distributions compared to the power-law Zone-I distributions is also consistent with the location of the mean-energy peak in the Zone-II aurora, as discussed previously. 
As can be seen by the relative perijove counts in each bin, the Zone-I and Zone-II spectra contain a similar amount of main-emission crossings in each hemisphere (though the average spectra of both zones typically contain more southern crossings, due to the lack of reliable northern crossings toward later perijoves), and thus the difference in the average Zone-I/II spectra is unlikely to be a sampling artefact.

\section{Discussion}
\label{sec:discussion}

This work has demonstrated that the peak in mean electron energy within the main emission is consistently found poleward of the peak in electron energy flux, and that this mean-energy peak is likely associated with the Zone-II of the main emission.
The precipitating energy-flux profile also has a significant contribution from the Zone-II or bridge aurora \citep{head+:2025} in the dusk sector, and thus the power output of the main emission cannot be said to be overwhelmingly dominated by either the Zone-I or Zone-II regions.
The poleward nature of the mean-energy peak has previously been noted by \citet{allegrini+:2020} based on JADE and JEDI observations during crossings of the main emission by Juno, though no explanation was given for this behaviour.
It is possible that the poleward nature of the mean-energy peak is a consequence of the lower ionospheric electron density in the poleward main emission \citep{kruegler+:2025}.
In both the Alfvénic (Equation \ref{eq:power_law_alfvenic}) and quasi-static-potential (Equation \ref{eq:power_law_currents}) cases, the same energy flux (which is essentially controlled by the strength of the acceleration process above the ionosphere) should lead to a higher mean precipitating-electron energy in the ionosphere in regions of lower density.
This idea is supported by the results shown in Figure \ref{fig:ebert2019_analogue}, in which the electron distributions above the Zone-II aurora consistently show profiles with greater electron fluxes at higher energy than those seen above the Zone-I aurora.
As indicated by Equations \ref{eq:power_law_alfvenic} and \ref{eq:power_law_currents}, different acceleration processes may have different dependencies on ionospheric electron density; unfortunately, the lack of reliable density estimates in the less-dense Zone-II aurora \citep{kruegler+:2025} prevents us from estimating the dependency of its flux-energy relation on density and hence from identifying whether the Zone-I and Zone-II have different origins.  
Alternatively, whereas the calculation of mean electron energy from UVS data supposed a uniform methane homopause altitude within the auroral region, it is possible that there exists a ``methane dome'' within the polar-cap aurora \citep{sinclair+:2020}.
This may lead to a greater absorption of lower-energy electrons (since they would have sufficient energy to penetrate deep enough to reach the raised methane homopause) than that which would be expected from a lower uniform methane homopause and hence lead to an overestimation of the mean electron energy inferred from UVS data.
However, the fact that the in-situ mean electron energy measured by JEDI (Figure \ref{fig:map_jedi_energy_proj}) is also consistently higher above the poleward main emission makes this scenario less necessary to explain the results of this work.

However, the different electron-spectrum behaviour at high energy flux between the two zones (power law for Zone-I, flattop for Zone-II), shown in Figure \ref{fig:ebert2019_analogue}, may indicate a phenomenological difference that is masked by the lack of density measurements in the Zone-II aurora.
These spectra may also explain why the mean-energy peak is located in the Zone-II aurora rather than more centrally in the main emission; at higher energy fluxes, the flattop Zone-II distributions have a greater contribution from 100-keV-scale electrons than the power-law Zone-I distributions.
In most UV-energy-flux ranges in Figure \ref{fig:ebert2019_analogue}, though especially beyond $10^{1.25}$ mW m$^{-2}$, the Zone-I electron flux is lower than the Zone-II electron flux; since the Zone-I and Zone-II necessarily have similar energy fluxes within the same bin, we interpret this as the presence of a greater proportion of lower-energy (< 20 keV) precipitating electrons in the Zone-I aurora required to produce energy fluxes comparable to the Zone-II aurora.
\citet{ebert+:2019} identified flattop distributions within the JADE energy range at weaker energy fluxes of $\sim$10-60 mW m$^{-2}$ just poleward of the main emission when Juno was passing over a bridge-like (Zone-II) auroral morphology during PJ5.
Our results suggest that these flattop distributions are very common in the Zone-II in the JEDI energy range at higher energy fluxes, and comparatively uncommon or absent in the Zone-I aurora.
\citet{mauk+:2017} identified an high downward electron energy flux ($\sim$280 mW m$^{-2}$) and flattop distribution (their Figure 5, B, E6) above the southern main emission during PJ1.
Given the presence of a visible bridge in PJ1-S and thus the confirmation of a considerably bright Zone-II main emission \citep{bonfond+:2017_pj1,head+:2025}, we anticipate that the profile seen by \citet{mauk+:2017} is essentially shaped by the contribution from the Zone-II electrons.
This suggested phenomenological difference between the Zone-I and Zone-II aurora is not entirely without precedent, as previous work has linked the Zone-II aurora with the presence of upward electron beams and thereby-induced large-amplitude electrostatic waves, features not associated with the Zone-I aurora \citep{elliott+:2020, sulaiman+:2022,head+:2025}.

Similarly, it is also notable that, over Juno's first 60 perijoves, inverted-V structures were only rarely seen in downward JEDI electron data (11\% of cases), compatible with previous work \citep{salveter+:2022}, and exclusively above the Zone-I aurora.
It is a source of ongoing debate to what extent corotation-breakdown currents and their ensuing quasi-static potentials contribute to the main auroral emission at Jupiter \citep{bonfond+:2020, provan+:2026}.
Indeed, the use of observed inverted-V structures in the electron energy spectra as a tracer for the presence of quasi-static potentials above Jupiter's ionosphere is not without its complications, insofar as corotation-breakdown currents may be able to induce high-latitude Alfvénic acceleration under certain plasma conditions \citep{sulaiman+:2026} and Alfvén waves may be able to induce discrete potential structures similar to inverted-Vs \citep{wu+:2025}.
Nevertheless, with our current understanding, the relative paucity of inverted-V structures above the main emission (even with the observational caveats discussed in section \ref{sec:observations} in mind) compared to signatures of broadband bidirectional acceleration may indicate that stochastic acceleration processes, such as those due to Alfvénic turbulence in the middle magnetosphere \citep{saur+:2003,saur+:2018}, dominate the power output of the main emission.
This case may be more strongly made for the Zone-II aurora in particular, due to their distinctly broadband electron energy spectra, especially at higher energy fluxes (Figure \ref{fig:ebert2019_analogue}), and the total absence of observed inverted-V structures.
It is presently unclear why occasional inverted-V structures would be seen exclusively above the Zone-I aurora.

We suggest that, when analysing the main emission, it may be necessary or advantageous to properly consider it as the combination of the Zone-I and Zone-II aurorae.
In this regard, the main emission has three important properties:
\begin{enumerate}
    \item It is separated into the upward-current Zone-I and downward-current Zone-II, both of which contribute significantly to the total main-emission power output \citep{mauk+:2020,head+:2025}.
    \item The Zone-I and Zone-II main emission may be morphologically indistinguishable within the aurora or may take the shape of distinct features (such as bridges in the dusk-side aurora during magnetospheric compression; \citealt{head+:2025}).
    \item It is unclear to what extent the Zone-I and Zone-II aurora are phenomenologically distinct.
\end{enumerate}
This latter property, regardless of whether the Zone-I and Zone-II aurora are phenomenologically linked or not, requires us to treat the main emission with some degree of subtlety.
If the Zone-I and Zone-II aurora are phenomenologically related (through a single source process, albeit one that may express itself differently in the aurora due to, for example, different ionospheric electron densities in the two zones), then bridges, which appear frequently in the dusk-side aurora, must necessarily be considered part of the main emission and be included when any analysis of the main emission is performed.
Historically, bridges (or ``parallel polar arcs'' or similar) have been frequently excluded from the main emission \citep[e.g.][]{nichols+:2009b,greathouse+:2021,palmaerts+:2023} though they often represent a significant contribution of the Zone-II aurora to the full main-emission power output.
If, instead, the Zone-I and Zone-II aurora originate from fundamentally different source processes, then they must be considered separately, even in cases where they are visually indistinguishable in the aurora (dawn-side main emission, dusk-side main emission without magnetospheric compression; \citealt{head+:2025}).
Otherwise, measured properties of the main emission (e.g. energy-flux relations, typical electron spectra) would not be identifiable as the consequence of a particular source process since the source processes of both the Zone-I and Zone-II aurora would contribute significantly.  
In this latter case, the term ``main emission'' would be solidified as an empirical descriptor similar to the ``polar emission'' (which encompasses a number of features with different accepted origins, such as ``bright polar spots''; \citealt{haewsantati+:2021}) instead of a phenomenological descriptor.
In this case, the Zone-I and Zone-II aurorae would be the phenomenological descriptors and it would be more sensible to refer to the ``main emissions'' rather than a single unified ``main emission''.
This work presents some evidence that the Zone-I and Zone-II aurorae may be phenomenologically distinct, though this certainly requires further work to be more robustly confirmed.

\section{Conclusions}
The main findings of this work are as follows:
\begin{enumerate}
    \item The peak in mean energy of precipitating electrons in Jupiter's UV main auroral emission is consistently found poleward of the energy-flux (brightness) peak over the first 60 perijoves.
    \item This mean-energy peak is likely associated with the Zone-II main emission.
    \item Inverted-V structures are seen infrequently in the downward electron spectra (11\% of cases) above the main emission during the first 60 perijoves and exclusively above the Zone-I aurora.
    \item The mean-electron-energy/energy-flux relationship everywhere within the main emission is approximately consistent with a power law of the form $\langle E \rangle=10^{A}\cdot F_E^{0.5}$, in line with theoretical predictions for both a current-based and Alfvénic main emission.
    \item The power-law scaling factor $A$ in the Zone-I aurora shows an inverse dependency with the ionospheric electron density broadly compatible with either a current-based or Alfvénic main emission.
    \item The Zone-I and Zone-II downward-electron energy spectra show distinct differences in behaviour, insofar as the Zone-II spectra transition to a clearly broadband ``flattop'' distribution at higher total energy flux. 
\end{enumerate}
Further reliable measurements of the ionospheric electron density in the low-density Zone-II main emission would allow us to identify whether the density dependence of the Zone-I aurora continues into the Zone-II aurora, or whether this latter zone has a different density dependence, which may support the idea that the two zones have different dominant source processes.
Further investigation of the main emission, with separate analysis for the two zones that constitute it, may also confirm whether the inverted-V structures, seen infrequently and exclusively above the Zone-I aurora in this work, are indeed exclusive to the Zone-I aurora and to what extent they are a necessary and sufficient indicator of acceleration by quasi-static potentials above the main emission.

\section*{Data availability}
Juno data can be obtained from the NASA Planetary Data System (\url{https://pds-atmospheres.nmsu.edu/data_and_services/atmospheres_data/JUNO/juno.html}).

\section*{Acknowledgements}
   We are grateful to NASA and contributing institutions that have made the Juno mission possible. This work was funded by NASA's New Frontiers Program for Juno via contract with the Southwest Research Institute. This publication benefits from the support of the French Community of Belgium in the context of the FRIA Doctoral Grant awarded to L. A. Head. The authors are grateful to G. Sicorello for his work in producing the main-emission masks.

\bibliography{references}

\begin{appendix}

\section{Derivation of power laws}
\label{app:power_law_derivation}
\renewcommand{\thefigure}{A.\arabic{figure}}
\renewcommand{\thetable}{A.\arabic{table}}
\renewcommand{\theequation}{A.\arabic{equation}}

\subsection{Alfvénic turbulence}
\label{sec:derivation_alfvenic}
From \citet{clark+:2018},
\begin{equation}
    F_E=\frac{\Sigma_A}{\lambda_e^2}\langle E \rangle^{2},
\end{equation}
where $F_E$ is the energy flux, $\langle E \rangle$ the mean energy of precipitating electrons, $\Sigma_A$ the effective Alfvénic conductance, and $\lambda_e$ the electron inertial length.
Since $\Sigma_A=(\mu_0 v_A)^{-1}$ and $\lambda_e = (\epsilon_0 m_e c^2 n_e^{-1} e^{-2})^{0.5}$, we can invert the expression to find
\begin{equation}
    \langle E \rangle=\sqrt{\frac{\epsilon_0 m_e c^2}{n_e e^2}}\mu_0^{0.5}v_A^{0.5} F_E^{0.5},
\end{equation}
where $\epsilon_0$ is the permittivity of free space, $m_e$ the electron mass, $c$ the speed of light, $n_e$ the electron number density, $e$ the elementary charge, $\mu_0$ the permeability of free space, and $v_A$ the Alfvén speed.
We use definition of the relativistic Alfvén speed \citep{lysak:2023}
\begin{equation}
    v_A=\frac{B}{\sqrt{\mu_0 m_i n_e + \frac{B^2}{c^2}}}
\end{equation} where $B$ is the magnetic field strength, $c$ the speed of light, and $m_i$ the ion mass, for a simple plasma ($n_i=n_e$), and see that the dependence of $v_A$ on electron density $n_e$ will vary between $v_A \propto n_e^{-0.5}$ in the non-relativistic case and $v_A \propto n_e^{0}$ in the fully relativistic case.
Thus,
\begin{equation}
    \langle E \rangle=\epsilon_0^{0.5} m_e^{0.5} m_i^{-0.25} c^{1} e^{-1} \mu_0^{0.25} B^{0.5} n_e^{[-0.75,-0.5]} F_E^{0.5}.
\end{equation}
Hence, for a power law of the form $\langle E \rangle=10^{A}F_E^k$, we would expect that $k=0.5$ and $A \propto n_e^{[-0.75,-0.5]}$ in the case of a main emission dominated by bidirectional acceleration originating from Alfvénic turbulence \citep{saur+:2018}.
Here, it is assumed that the system is dominated entirely by small-scale fluctuations and that all the energy of the Alfvén waves is used to produce the aurora \citep{clark+:2018}.
Further discussion of the assumptions made in this model is available in \citet{saur+:2003}.

\subsection{Currents/quasi-static potentials}

From \citet{clark+:2018},
\begin{equation}
    F_E = F_{0}\left[1+\left(\frac{e\langle E \rangle}{W_{th}}\right) + 0.5 \left(\frac{e\langle E \rangle}{W_{th}}\right)^2\right],
\end{equation}
where $W_{th}$ is the average thermal energy of precipitating electrons, $F_{0}$ is the precipitating flux of electrons in the loss cone in the absence of an accelerating potential \citep{cowley+bunce:2001}, defined as
\begin{equation}
    F_{0}=2 n_e W_{th}\left( \frac{W_{th}}{2 \pi m_e} \right)^{0.5},
\end{equation}
and other definitions are as per section \ref{sec:derivation_alfvenic}.
This equation can be expressed in a quadratic form as
\begin{equation}
    \frac{1}{2} F_{0} \left(\frac{e\langle E \rangle}{W_{th}}\right)^2 + F_{0}\left(\frac{e\langle E \rangle}{W_{th}}\right) + (F_0-F_E)=0
\end{equation}
and hence its solutions are given by 
\begin{equation}
    \frac{e\langle E \rangle}{W_{th}} = \frac{-F_{0}\pm\sqrt{F_0^2-2F_0(F_0-F_E) }}{F_0 }
\end{equation}
or equivalently
\begin{equation}
    \frac{e\langle E \rangle}{W_{th}} = -1 \pm \sqrt{ \frac{2F_E}{F_{0}} - 1 }.
\end{equation}
If we assume that the contribution to the auroral energy flux from the quasi-static potentials dominates over the ``baseline'' flux from loss-cone electrons, and hence $F_E \gg F_{0}$,
\begin{equation}
    \frac{e\langle E \rangle}{W_{th}} = 2^{0.5}F_{0}^{-0.5} F_E^{0.5},
\end{equation}
and hence, substituting for $F_{0}$,
\begin{equation}
    \langle E \rangle=e^{-1} W_{th}^{0.25} 2^{0.25} \pi^{0.25} m_e^{0.25} n_e^{-0.5} F_E^{0.5} 
\end{equation}
Hence, for a power law of the form $\langle E \rangle=10^{A}F_E^k$, we would expect that $k=0.5$ and $A \propto n_e^{-0.5}$ in the case of a main emission dominated by monoenergetic acceleration originating from quasi-static potentials.

This derivation uses the expression from \citet{clark+:2018} for non-relativistic electrons accelerated over an infinitely tall electrical potential.
The assumption of electron non-relativity is considered a reasonable simplification, since main-emission crossings have typical average electron energies of 10-100 keV (see Figure \ref{fig:map_uv_energy_proj}), which is below the electron rest energy of 511 keV.
As can be seen in Figure 4b of \citet{clark+:2018}, the non-relativistic and relativistic ``$R_B \rightarrow \infty$'' relations remain very similar across the relevant energy range.
Indeed, the assumption that the acceleration occur over an infinitely tall potential structure (or rather that the magnetic field strength at the top of the potential structure is zero) seems to play a larger role than relativistic effects, though not enough to meaningfully affect the conclusions of this work.
Further discussion of the assumptions made in this model is available in \citet{cowley+bunce:2001}.


\section{Zone-I and II inversion: PJ14-S and PJ18-S}
\label{app:z12_south_inversion}
\renewcommand{\thefigure}{B.\arabic{figure}}
\renewcommand{\thetable}{B.\arabic{table}}

\begin{figure}[tb]
    \centering
    \includegraphics[width=0.6\linewidth]{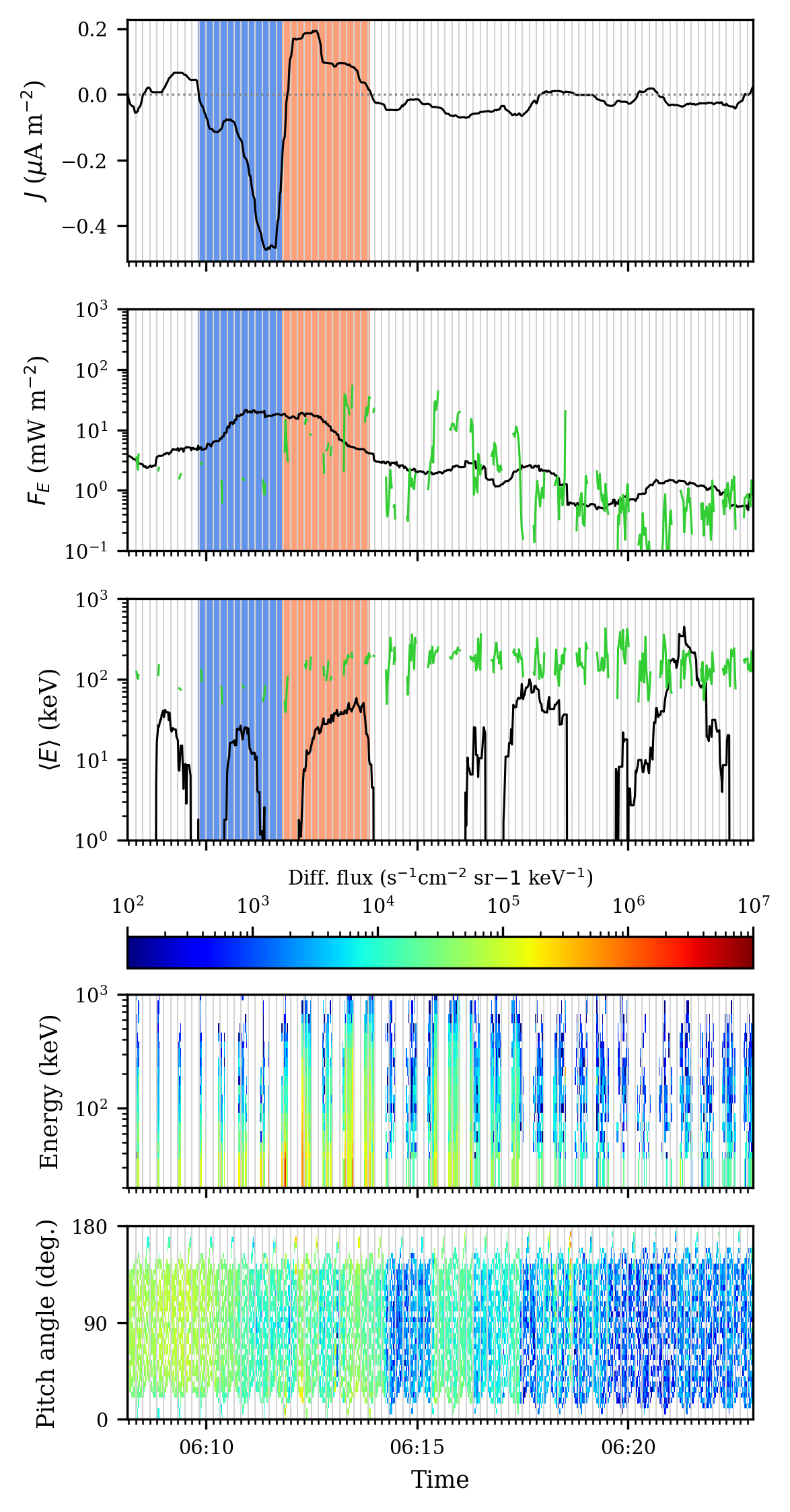}
    \caption{
        Measured and derived parameters using the PJ14-S main emission crossing. From top to bottom: field-aligned current density (positive = upward); energy flux (black = UVS, green = JEDI); mean downward electron energy; JEDI downward energy spectrum; JEDI pitch angle spectrum. Red and blue rectangles denote the designated Zone-I and Zone-II crossings.
    }
    \label{fig:pj14s_inversion}
\end{figure}

\begin{figure}[tb]
    \centering
    \includegraphics[width=\linewidth]{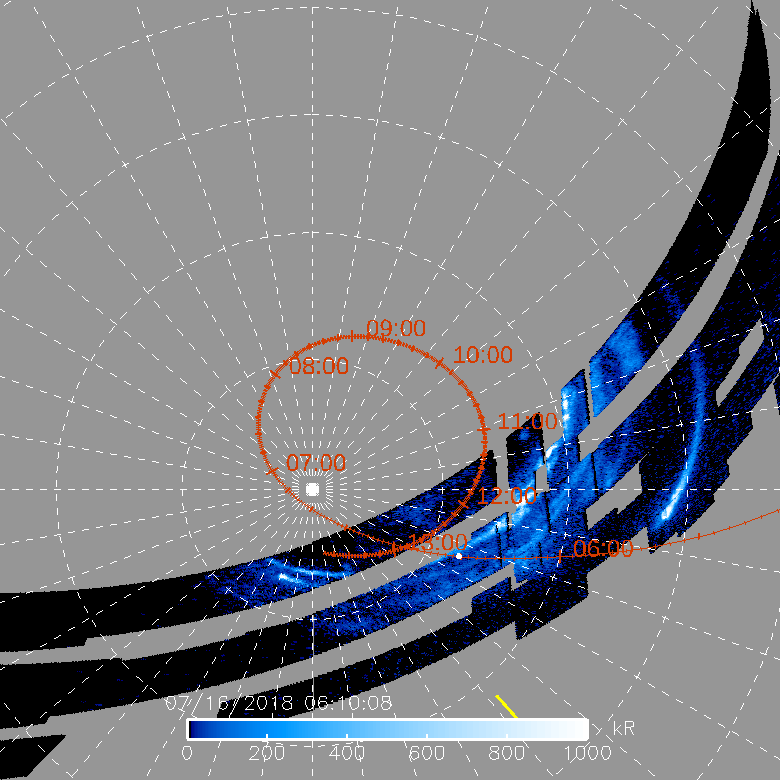}
    \caption{
        Instantaneous auroral energy flux inferred from UVS during PJ14-S at 06:10. Gridlines are spaced by 10\textdegree. The Juno footpath is given in red, Juno's footprint by the white dot, and subsolar longitude in yellow.
    }
    \label{fig:pj14s_before}
\end{figure}

\begin{figure}[tb]
    \centering
    \includegraphics[width=\linewidth]{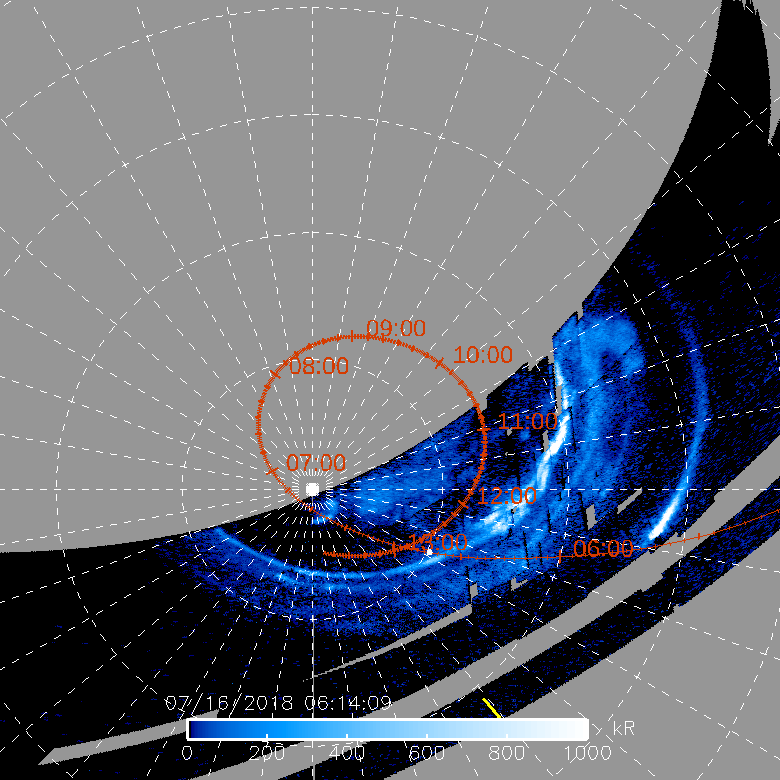}
    \caption{
        As Figure \ref{fig:pj14s_before} but four minutes later at 06:14.
    }
    \label{fig:pj14s_after}
\end{figure}

\begin{figure}[tb]
    \centering
    \includegraphics[width=0.6\linewidth]{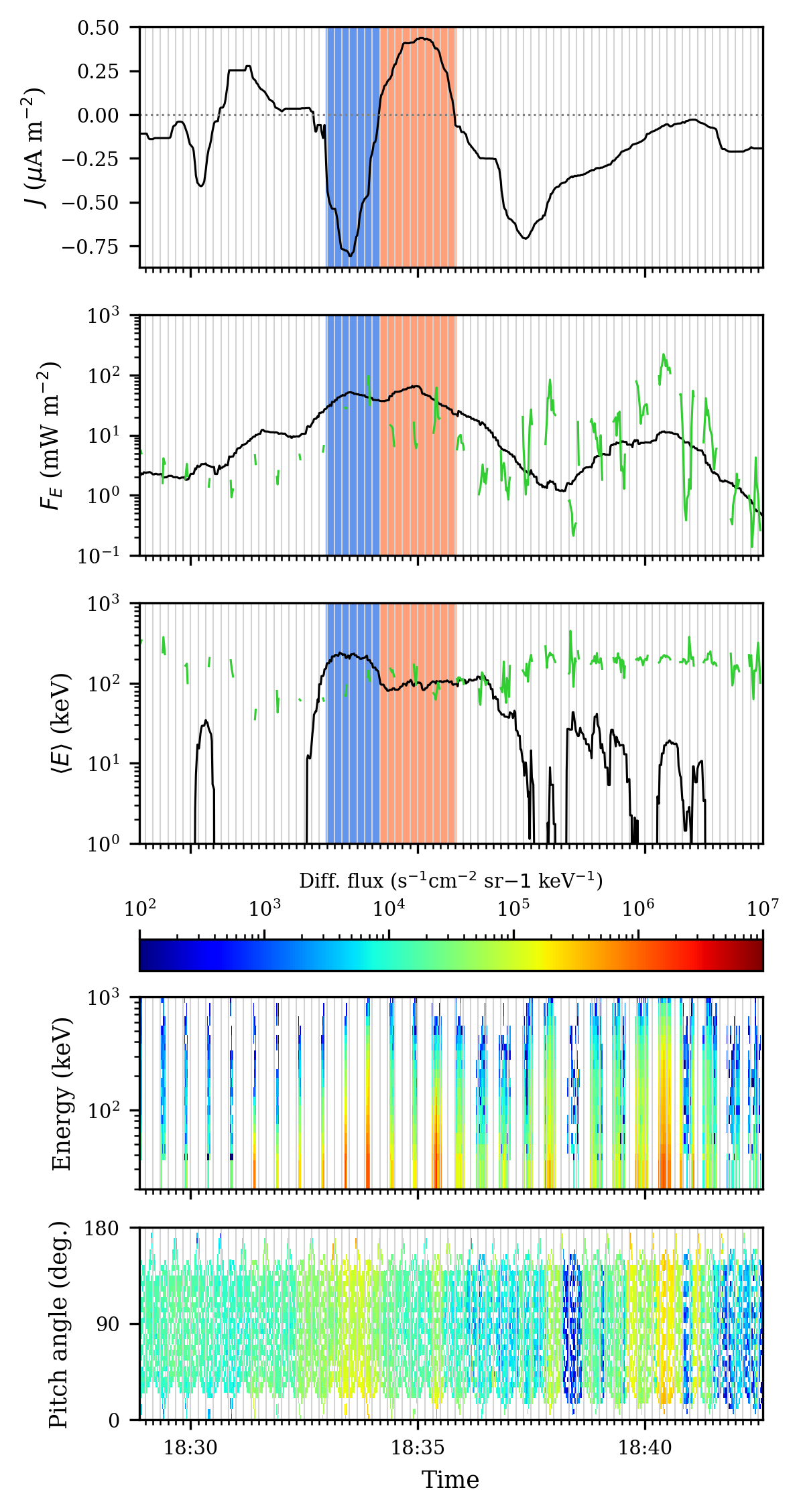}
    \caption{
        Measured and derived parameters using the PJ18-S main emission crossing, as per Figure \ref{fig:pj14s_inversion}.
    }
    \label{fig:pj18s_inversion}
\end{figure}

\begin{figure}[tb]
    \centering
    \includegraphics[width=\linewidth]{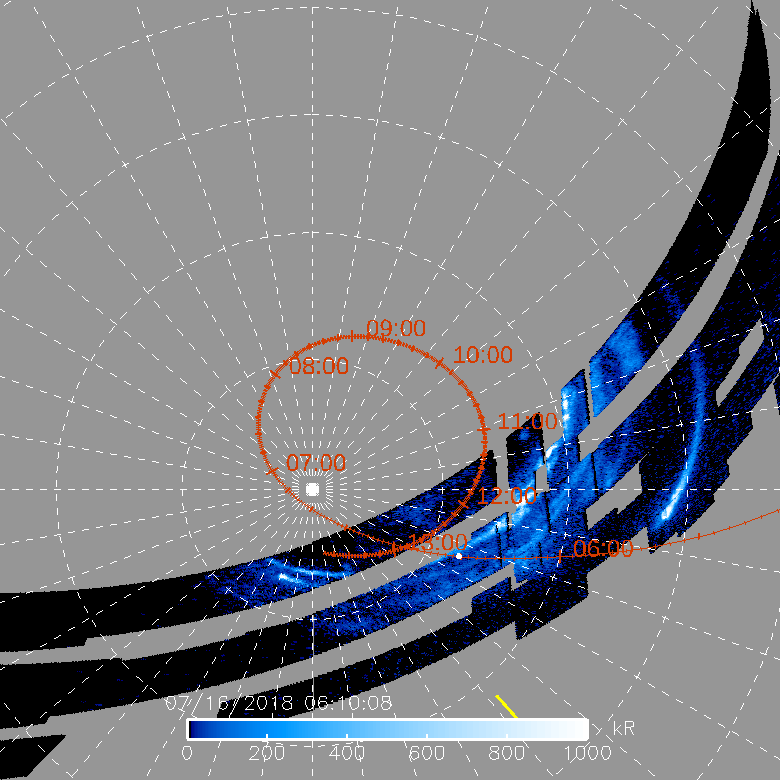}
    \caption{
        Instantaneous auroral energy flux inferred from UVS during PJ18-S at 18:33. Gridlines are spaced by 10\textdegree. The Juno footpath is given in red, Juno's footprint by the white dot, and subsolar longitude in yellow.
    }
    \label{fig:pj18s_before}
\end{figure}

\begin{figure}[tb]
    \centering
    \includegraphics[width=\linewidth]{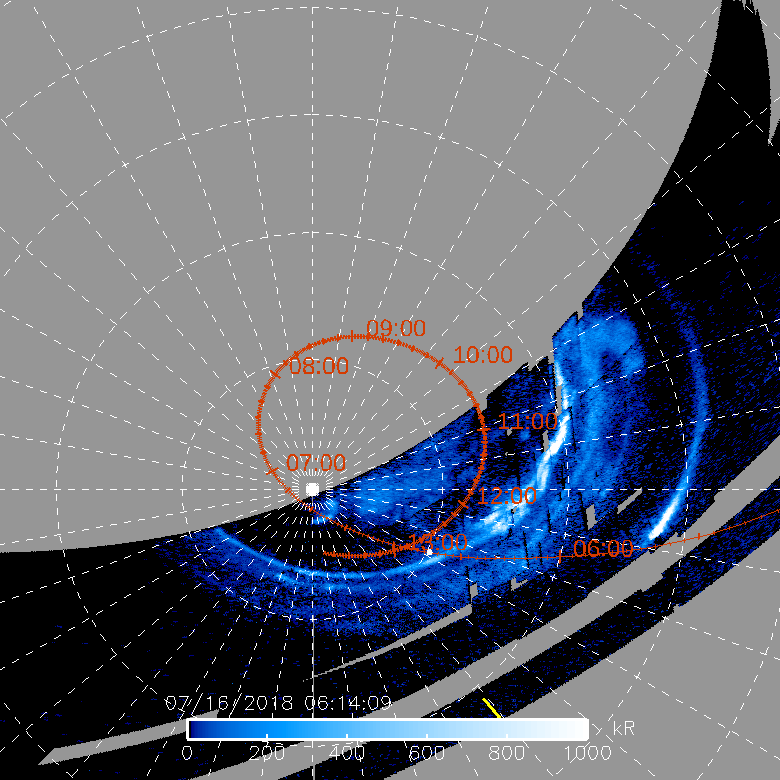}
    \caption{
        As Figure \ref{fig:pj18s_before} but three minutes later at 18:36.
    }
    \label{fig:pj18s_after}
\end{figure}

Figures \ref{fig:pj14s_inversion} and \ref{fig:pj18s_inversion} show the time series of various relevant parameters during Juno crossings of the main emission for PJ14-S and PJ18-S respectively.
In both cases, the Juno footprint crosses the main emission from the outside in the poleward direction; in the typical configuration, we would therefore expect the Zone-I aurora to be crossed first, followed by the Zone-II aurora.
Instead, in PJ14-S especially, the Zone-II crossing clearly precedes the Zone-I crossing; both zones show the correct current direction and contain enhanced broadband downward electron flux typical of the main emission \citep{mauk+:2020}.
Additionally, the mapped Juno footprint was clearly passing through the UV main emission during this period, as can be seen in Figures \ref{fig:pj14s_before} and \ref{fig:pj14s_after}.
The case is similar for PJ18-S, though here the Zone-I crossing is succeeded by a negative current region of comparable intensity to the designated Zone-II aurora.
This second negative-current region may correspond to the crossings of the bridge, which may suggest that the Zone-I aurora may sometimes find itself between or within the Zone-II aurora.
Alternatively, based on the approximately ``butterfly'' distribution of the electrons during the designated Zone-II crossing, this may actually represent the crossing of an injection signature on or near the main emission \citep{head+:2026}, which does have a noticeably more disrupted morphology (Figures \ref{fig:pj18s_before} and \ref{fig:pj18s_after}) compared to the PJ14-S crossing, with the second negative-current region as the ``true'' Zone-II aurora.
In all, we consider these two crossings to represent cases with potentially inverted Zone-I/II aurorae in the southern hemisphere, completing the set of four possibilities (typical/inverted, north/south) proposed by \citet{al_saati+:2022}.

\section{Choice of maximum pitch angle}
\label{app:pitch_angle_check}
\renewcommand{\thefigure}{C.\arabic{figure}}
\renewcommand{\thetable}{C.\arabic{table}}

\begin{figure}[tb]
    \centering
    \includegraphics[width=\linewidth]{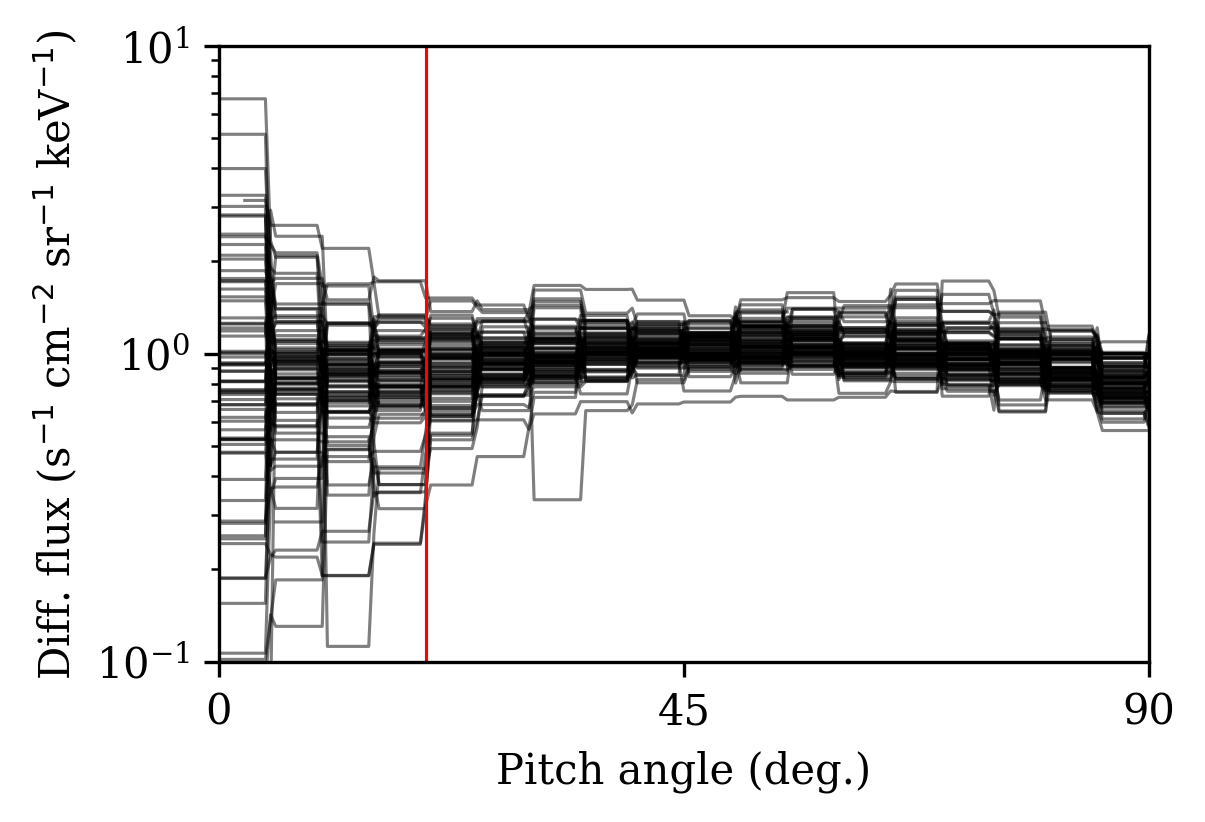}
    \caption{
        Normalised pitch-angle distribution of downward electrons during Juno crossings of the main emission from PJ1 to PJ60. The cutoff pitch angle (20\textdegree) is given by the red line. Pitch angles in the northern hemisphere have been reversed to ensure sensible comparison between the two hemispheres (0\textdegree\ = toward the planet).
    }
    \label{fig:pitch_angle_check}
\end{figure}

Since main-auroral acceleration in both the current-based and Alfvénic scenarios is expected to be field-aligned, the choice of cutoff pitch angle for ``downward'' (or indeed upward) electrons in this work must contain the field-aligned portion but not necessarily perfectly match the loss-cone angle at all altitudes.
This is because, in an idealised scenario, the auroral electron population would be perfectly aligned with the magnetic field vector and there would be no background electron flux outside of this direction.
The normalised (profile - profile median) pitch-angle distributions for the main-emission crossings considered in this work are shown in Figure \ref{fig:pitch_angle_check}.
It is clear that the largest proportion of the field-aligned electron flux is found within 20 \textdegree\ of the local magnetic field vector.
It is likely the the profiles showing reductions in field-aligned electron flux, instead of enhancements, are due to the influence of injection signatures near the main emission and their ``butterfly'' pitch-angle distributions \citep{head+:2026}.
In any case, we consider that a cutoff angle of 20\textdegree\ suitably captures the field-aligned portion of the downward electron flux.

\section{Altitude dependence of JEDI/UVS energy-flux ratio}
\label{app:uvs_jedi_flux}
\renewcommand{\thefigure}{D.\arabic{figure}}
\renewcommand{\thetable}{D.\arabic{table}}

\begin{figure}[tb]
    \centering
    \includegraphics[width=\linewidth]{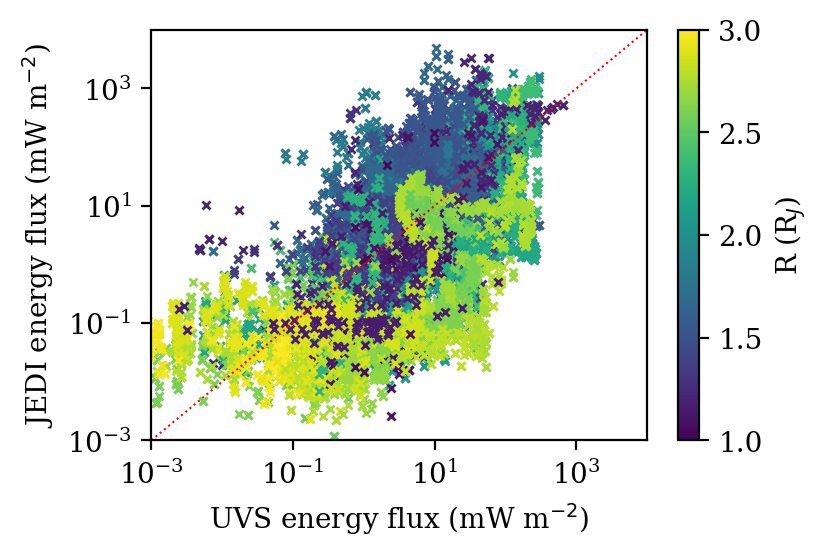}
    \caption{
        Auroral energy flux inferred from UVS against downward-electron energy flux measured by JEDI for every timestamp in every main-emission crossing considered in this work. Juno's instantaneous distance from the centre of Jupiter $R$ is denoted by the colour of each data point. The 1:1-ratio line is given in red. 
    }
    \label{fig:uvs_jedi_flux}
\end{figure}

In an ideal scenario, the energy flux inferred from UVS observations of the aurora and JEDI observations of the downward-electron flux should be identical.
As shown in Figure \ref{fig:uvs_jedi_flux}, the distribution of UVS and JEDI energy fluxes is broadly consistent with the expected 1:1 relation, albeit with a significant degree of scatter (approximately one order of magnitude above and two orders of magnitude below) around this ideal relation.
It is noteworthy that there appears to be some measure of altitude dependence of this deviation from the ideal relation; main-emission crossings at altitudes of $\sim$1.25-2 R$_{J}$ consistently measure lower in-situ electron energy fluxes than that which would be expected from the UV auroral energy flux.
We suggest that this is consistent with the existence of a broad auroral acceleration region below an altitude of 2 R$_{J}$.
While some measure of auroral acceleration will occur between 1 and 2 R$_{J}$, most is expected to occur below 1 R$_{J}$ \citep{hess+:2010,sulaiman+:2022}, which would mean that a significant portion of the downward electron acceleration would occur below Juno at higher altitudes and thus not be detected by JEDI.

\section{Case-by-case density comparison}
\label{app:density_cases}
\renewcommand{\thefigure}{E.\arabic{figure}}
\renewcommand{\thetable}{E.\arabic{table}}

\begin{figure}[tb]
    \centering
    \includegraphics[width=\linewidth]{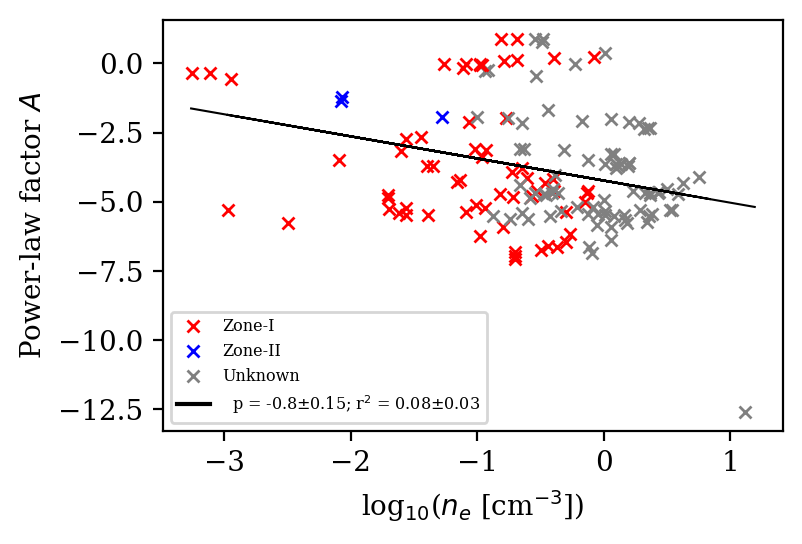}
    \caption{
        Ionospheric electron density $n_e$ against power-law factor $A$ measured during Juno crossings of the main emission. Data when Juno was passing over an identified Zone-I or II aurora are given in red and blue, respectively, with grey points denoting data that could not be reliably attributed to either zone. The fitted relation is given by a solid black line. 
    }
    \label{fig:density_casebycase}
\end{figure}

To complement the statistical results presented in Figure \ref{fig:A_vs_density}, we additionally investigated the relation between ionospheric electron density and power-law factor $A$ on a case-by-case basis.
Here, for each point during a Juno crossing of the main emission with calculated density values \citep{kruegler+:2025}, a power law (as per Equation \ref{eq:power_law}) was fit to the energy-flux and mean-energy values present in a 2\textdegree-$\times$-2\textdegree\ latitude-longitude region around the Juno footprint in the perijove auroral maps (see section \ref{sec:observations}).
Fitted power laws with R$^{2}$ values below 0.5 were rejected to ensure that only reasonably reliable power laws were considered.
Figure \ref{fig:density_casebycase} shows a very weak negative relation between ionospheric electron density and power-law factor $A$.
Interestingly, the density dependence $p=-0.8\pm0.15$ is compatible with the $p$ derived from Figure \ref{fig:A_vs_density}; however, due to the large scatter in the points (and the appropriately weak R$^{2}$), this cannot be said to be a confirmation of the results of Figure \ref{fig:A_vs_density}.
As mentioned previously, we interpret the weakness of the relation in Figure \ref{fig:density_casebycase} and the comparative strength of the relation in Figure \ref{fig:A_vs_density} as an indication that, while ionospheric electron density influences the properties of the main emission in a general sense, other effects may dominate over the density dependence during a given perijove.

\section{Crossings considered in this work}
\label{app:crossing_table}
\renewcommand{\thefigure}{F.\arabic{figure}}
\renewcommand{\thetable}{F.\arabic{table}}

\setlength\tabcolsep{0.07cm}
\centering
\tiny
\topcaption{Juno main-emission crossings used in this work.}
\label{tab:crossing_table}
\tablehead{\toprule PJ & HS & Zone-I & Zone-II & $\land$? & $\leftrightharpoons$? & Exp.\\ \midrule}
\tabletail{\midrule \multicolumn{7}{l}{Continued...} \\ \bottomrule}
\tablelasttail{\midrule \multicolumn{7}{l}{ \parbox{\columnwidth}{`PJ'' gives the perijove number. ``HS'' refers to the hemisphere of the Juno crossing. ``Zone-\#'' gives the UTC timestamps of the start and end of the Zone-I and Zone-II crossing. ``$\land$?'' indicates whether an inverted-V structure was seen in the JEDI data above the Zone-I aurora. Where ``$\land$?'' is ``Yes'' with no Zone-I timestamps, this indicates that an inverted-V structure was seen over a region of the main emission that could conceivably be part of the Zone-I aurora but where this could not be reliably determined. No cases were seen of an inverted-V structure above the Zone-II aurora. ``$\leftrightharpoons$?'' denotes inverted cases where the Zone-I aurora was determined to be poleward of the Zone-II aurora. ``Exp.'' denotes the main-emission expansion in km (uncertainty $\sim$ 300 km) from the reference contours given in \citet{head+:2024}. Negative expansions denote contractions of the main emission from its average position.}} \\ \bottomrule}

\begin{supertabular}{rlllllr}

1     & N     & --    & --    & --    & --    & 100 \\
1     & S     & 13:31:16-13:33:11 & 13:33:11-13:40:00 & --    & --    & 700 \\
3     & N     & 16:23:25-16:24:10 & 16:22:50-16:23:25 & --    & --    & 1300 \\
3     & S     & 17:36:50-17:37:42 & 17:37:42-17:38:30 & --    & --    & 300 \\
4     & N     & 12:24:51-12:26:00 & 12:23:30-12:24:51 & --    & --    & -600 \\
4     & S     & 13:37:40-13:39:15 & 13:39:15-13:40:20 & Yes   & --    & -100 \\
5     & N     & 08:29:00-08:30:00 & 08:28:00-08:29:00 & --    & --    & -1700 \\
5     & S     & 09:42:40-09:45:10 & 09:45:10-09:48:50 & --    & --    & -800 \\
6     & N     & 05:28:00-05:28:54 & 05:28:54-05:29:40 & --    & Yes   & -400 \\
6     & S     & 06:52:00-06:53:10 & 06:53:10-06:59:30 & --    & --    & -300 \\
7     & N     & 01:15:20-01:16:20 & 01:14:10-01:15:20 & Yes   & --    & -400 \\
7     & S     & 02:32:30-02:34:10 & 02:34:10-02:34:50 & --    & --    & 0 \\
8     & N     & --    & --    & --    & --    & 600 \\
8     & S     & 22:27:20-22:28:02 & 22:28:02-22:28:50 & --    & --    & 400 \\
9     & N     & 17:18:40-17:19:40 & 17:17:00-17:18:40 & --    & --    & -600 \\
9     & S     & 18:32:50-18:35:40 & 18:35:40-18:40:20 & --    & --    & -300 \\
10    & N     & 17:24:00-17:25:00 & 17:25:00-17:25:30 & Yes   & Yes   & -2100 \\
10    & S     & --    & --    & --    & --    & -900 \\
11    & N     & 13:32:05-13:32:30 & 13:31:30-13:32:05 & --    & --    & 300 \\
11    & S     & 14:46:10-14:47:20 & 14:47:20-14:48:30 & --    & --    & 800 \\
12    & S     & 10:44:50-10:48:20 & 10:48:20-10:52:30 & --    & --    & -1000 \\
13    & N     & 05:04:30-05:05:22 & 05:05:22-05:06:30 & --    & Yes   & 600 \\
13    & S     & --    & --    & --    & --    & 900 \\
14    & N     & 04:40:45-04:41:25 & 04:41:25-04:42:20 & Yes   & Yes   & 200 \\
14    & S     & 06:11:50-06:13:50 & 06:09:50-06:11:50 & --    & Yes   & 500 \\
15    & S     & 01:55:50-01:58:20 & 01:58:20-02:02:30 & --    & --    & -500 \\
16    & N     & --    & --    & --    & --    & 400 \\
16    & S     & 21:56:40-21:59:30 & 21:59:30-22:01:10 & --    & --    & 400 \\
17    & N     & --    & --    & --    & --    & -400 \\
17    & S     & --    & --    & --    & --    & 200 \\
18    & N     & 17:14:20-17:15:00 & 17:13:30-17:14:20 & --    & --    & -- \\
18    & S     & 18:34:10-18:35:50 & 18:33:00-18:34:10 & --    & Yes   & -700 \\
19    & N     & --    & --    & --    & --    & 200 \\
19    & S     & --    & --    & --    & --    & 700 \\
20    & N     & --    & --    & --    & --    & -200 \\
20    & S     & --    & --    & --    & --    & 100 \\
21    & N     & 03:37:50-03:38:25 & 03:36:30-03:37:50 & --    & --    & -700 \\
21    & S     & 04:58:10-04:59:50 & 04:59:50-05:01:50 & --    & --    & -400 \\
22    & N     & 03:13:00-03:13:35 & 03:13:35-03:14:30 & --    & Yes   & 1800 \\
22    & S     & --    & --    & --    & --    & 1800 \\
23    & N     & 21:54:10-21:55:40 & 22:00:00-22:03:00 & --    & Yes   & 0 \\
23    & S     & --    & --    & --    & --    & 100 \\
24    & S     & --    & --    & --    & --    & 600 \\
25    & S     & --    & --    & --    & --    & 400 \\
26    & N     & --    & --    & --    & --    & -1600 \\
27    & N     & --    & --    & --    & --    & -900 \\
27    & S     & 11:14:10-11:16:00 & 11:16:00-11:17:30 & Yes   & --    & -200 \\
28    & N     & --    & --    & --    & --    & 2300 \\
28    & S     & --    & --    & --    & --    & 2100 \\
29    & N     & --    & --    & --    & --    & -1500 \\
29    & S     & 03:28:57-03:31:18 & 03:31:18-03:41:27 & --    & --    & -300 \\
30    & N     & --    & --    & --    & --    & 3000 \\
30    & S     & --    & --    & --    & --    & 1000 \\
32    & N     & 17:08:55-17:09:50 & 17:08:00-17:08:55 & --    & --    & -900 \\
32    & S     & 18:42:33-18:51:48 & 18:51:48-18:58:35 & --    & --    & -1000 \\
33    & N     & --    & --    & --    & --    & 0 \\
33    & S     & 00:29:25-00:34:37 & 00:34:37-00:42:35 & Yes   & --    & -500 \\
34    & N     & --    & --    & --    & --    & 800 \\
34    & S     & 08:42:04-08:44:10 & 08:44:10-08:55:44 & --    & --    & 300 \\
35    & N     & 07:27:20-07:28:22 & 07:28:22-07:35:50 & --    & Yes   & 600 \\
35    & S     & --    & --    & --    & --    & 100 \\
36    & N     & --    & --    & --    & --    & 300 \\
36    & S     & 00:07:10-00:11:11 & 00:11:11-00:15:00 & --    & --    & 900 \\
37    & N     & 16:25:28-16:26:26 & 16:26:26-16:32:00 & --    & Yes   & -200 \\
37    & S     & 18:24:18-18:28:19 & 18:28:19-18:33:30 & --    & --    & 0 \\
38    & N     & --    & --    & --    & --    & -200 \\
38    & S     & 15:36:00-15:39:10 & 15:39:10-15:41:00 & --    & --    & 300 \\
39    & N     & --    & --    & --    & --    & -1100 \\
39    & S     & 11:58:33-12:00:37 & 12:00:37-12:10:00 & --    & --    & 500 \\
40    & N     & 01:24:11-01:25:59 & 01:25:59-01:35:12 & --    & Yes   & -1500 \\
41    & N     & --    & --    & --    & --    & 200 \\
41    & S     & --    & --    & --    & --    & 400 \\
42    & N     & --    & --    & --    & --    & 700 \\
42    & S     & --    & --    & --    & --    & -300 \\
43    & S     & 10:21:20-10:23:30 & 10:23:30-10:24:40 & Yes   & --    & -1600 \\
45    & N     & --    & --    & --    & --    & -- \\
45    & S     & 18:29:55-18:31:32 & 18:31:32-18:39:00 & --    & --    & 300 \\
46    & N     & 20:50:23-20:51:33 & 20:51:33-05:59:50 & --    & Yes   & 1300 \\
46    & S     & --    & --    & --    & --    & 1000 \\
47    & N     & --    & --    & --    & --    & 800 \\
48    & S     & --    & --    & --    & --    & -600 \\
50    & S     & 09:42:11-09:42:54 & 09:42:54-09:52:00 & --    & --    & -200 \\
52    & N     & --    & --    & --    & --    & -- \\
53    & N     & --    & --    & --    & --    & -- \\
54    & N     & --    & --    & Yes   & --    & -- \\
55    & N     & --    & --    & --    & --    & -- \\
56    & N     & --    & --    & --    & --    & -1900 \\
58    & N     & --    & --    & --    & --    & -900 \\
58    & S     & 23:11:00-23:12:20 & 23:12:20-23:13:50 & Yes   & --    & -100 \\
59    & N     & --    & --    & --    & --    & 1400 \\
59    & S     & --    & --    & --    & --    & 1800 \\
60    & N     & --    & --    & Yes   & --    & -100 \\

\end{supertabular}

\end{appendix}

\label{LastPage}

\end{document}